\documentclass[aps,prb,reprint,showpacs,amsmath,amssymb]{revtex4-2}
\usepackage{newtxtext}
\usepackage{newtxmath}
\usepackage{graphicx}
\usepackage{subcaption}
\usepackage{braket}
\usepackage{bm}
\usepackage{siunitx}
\usepackage{xcolor}
\usepackage{hyperref} 
\newcommand{\nc}{\newcommand}
\nc{\red}[1]{\textcolor{red}{#1}} 
\nc{\tb}{\textbf}                 
\nc{\di}{\displaystyle}           
\nc{\lr}[1]{\left( #1 \right)}
\nc{\alr}[1]{\langle #1 \rangle}
\nc{\blr}[1]{\left[ #1 \right]}
\nc{\clr}[1]{\left\{ #1 \right\}}

\begin{document}

\title{Radiative coupling between plasmon and electron-hole pairs in \\ a metallic film based on extended Bohm-Pines theory }

\author{
Soshun Inoue\textsuperscript{1},
Takeshi Inaoka\textsuperscript{2},
Hajime Ishihara\textsuperscript{3}
}

\affiliation{\textsuperscript{1}Department of Material Engineering Science, Graduate School of Engineering Science, The University of Osaka, Toyonaka, Osaka 560-8531, Japan}
\affiliation{\textsuperscript{2}Department of Physics and Earth Science, Faculty of Science, University of the Ryukyus, 1 Senbaru, Nishihara-cho, Okinawa 903-0213, Japan}
\affiliation{\textsuperscript{3}Ritsumeikan Semiconductor Application Research Center (RISA), Ritsumeikan University, Kusatsu, Shiga 525-8577, Japan}

\date{\today}

\begin{abstract}
Hot carrier generation in metals, where high-energy electron-hole pairs are produced
via plasmon excitation, has emerged as a promising mechanism for photoelectric
conversion and photocatalysis. However, conventional theories often describe this
process through phenomenological relaxation via Landau damping, which fails to
account for the microscopic origin of the frequency-dependent internal quantum
efficiency (IQE) observed in experiments. To address this gap, we develop an extended
Bohm-Pines theory for a metallic thin film that explicitly incorporates light-matter
interactions within a non-local response framework. Our approach treats collective
(plasmonic) and individual (electron-hole) excitations on equal footing and includes
their coupling mediated by both longitudinal and transverse electromagnetic fields.
This results in a self-consistent theory of the optical response of metallic films.
The derived total Hamiltonian includes radiative corrections that recover the known
dispersion of surface plasmon polaritons and, importantly, predict a frequency-dependent
radiative coupling between collective and individual modes. This previously neglected
transverse coupling naturally explains the IQE peak near the plasmon resonance and
reveals a new mechanism of hot carrier generation distinct from conventional Landau
damping. Our results provide a unified theoretical foundation for understanding
plasmon-induced hot carrier dynamics and offer guidance for resonance-based photonic
design strategies to enhance energy conversion efficiency in metal nanostructures.
\end{abstract}

\maketitle

\onecolumngrid
\section{Introduction}
Elementary excitations in metals can be broadly classified into two categories: collective excitations, known as plasmons, and individual excitations, namely electron–hole (e–h) pairs. It is well established that optical excitation of surface plasmons at a metal interface can generate high-energy individual excitations—referred to as hot carriers—that lie beyond thermal equilibrium~\cite{brongersma2015plasmon}. The injection of such hot carriers into adjacent semiconductors or molecules enables photoelectric conversion across the visible to near-infrared spectrum, a process traditionally regarded as challenging. Consequently, hot carriers are being explored for various applications, including compact photodetectors that operate without complex optical components~\cite{li2015circularly}, and photocatalysts that facilitate key chemical reactions relevant to a decarbonized society, such as H$_2$O splitting and CO$_2$ reduction~\cite{ueno2016plasmon}. Despite these promising directions, the conversion efficiencies achieved thus far remain suboptimal.
To realize highly efficient photoelectric conversion, it is crucial to rapidly transfer energy from plasmons to hot carriers. However, conventional theories typically describe hot carrier generation as a phenomenological relaxation process via Landau damping~\cite{sundararaman2014theoretical,manjavacas2014plasmon,besteiro2017understanding,zhang2021theory}, leaving the microscopic mechanisms insufficiently explained. Recent experimental studies have revealed that the internal quantum efficiency (IQE)—which quantifies photoelectric conversion efficiency normalized by the amount of absorbed photons—exhibits a pronounced peak near the plasmon resonance energy~\cite{shi2018enhanced}. This behavior cannot be accounted for by relaxation processes alone, underscoring the necessity for a more comprehensive microscopic understanding.
Bohm and Pines~\cite{bohm1951collective1,pines1952collective2,bohm1953collective3,pines1953collective4} formulated a theoretical framework in which the electronic states in a metal are expanded in terms of both collective and individual excitation bases, thereby providing a means to discuss energy transfer between them. Notably, Yamada~\cite{yamada1960mechanism} demonstrated that within this model, longitudinal interactions between these excitations yield the Landau damping rate initially introduced by Landau. However, such treatments are inherently confined to bulk systems. For understanding hot carrier generation, it is imperative to extend the Bohm–Pines theory—originally developed for infinite bulk metals—to finite systems that allow for coupling with electromagnetic radiation.

In this study, we develop a generalized theoretical framework for the optical response of metals by considering an infinitely extended metallic thin film, explicitly incorporating light–matter interactions in this finite geometry. The proposed theory not only reproduces the conventional Drude response—where individual excitations are neglected—but also uncovers a previously unaccounted transverse interaction between collective and individual excitations, mediated by the radiation field. This radiative coupling has not been included in standard theories of hot carrier generation, yet it holds promise for resolving long-standing issues, such as the frequency-dependent behavior of IQE, and may offer a new route to enhance hot carrier generation efficiency.

This paper provides a detailed derivation of the extended Bohm–Pines theory, demonstrates the emergence of the well-known surface plasmon polariton dispersion relation~\cite{economou1969surface}, and discusses the novel effects anticipated from the radiative coupling.

\section{ Hamiltonian of matter}
\subsection{ Derivation of Bohm–Pines Hamiltonian in a metallic Film }
We consider a metallic film of thickness $d$, infinitely extended in the $x$–$y$ plane. The matter Hamiltonian of this system, $\hat{H}_{\mathrm{mf}}$, is described using an infinite quantum well potential $V(z)$ for electrons, based on the jellium model~\cite{fetter2012quantum}, as follows:
\begin{align}
  \hat{H}_{\mathrm{mf}} &= \sum_{i=1}^N \frac{\hat{\bm{p}}_i^2}{2m^{\ast}} + \frac{1}{2}\sum_{i\neq j}^{N} \frac{e^2}{4\pi\varepsilon_0 |\hat{\bm{r}}_i-\hat{\bm{r}}_j|} + V(z)
\label{H_mf} \\
  V(z) &= 
  \left\{
    \begin{matrix}
      0 \quad & (0 \leqq z\leqq d) \\
      \infty \quad & (z<0 \text{ or } z>d)
    \end{matrix}
  \right.
\label{well-potential}
\end{align}
Here, $N$ and $m^{\ast}$ denote the number and effective mass of electrons, respectively, and $\hat{\bm{p}}_i$ and $\hat{\bm{r}}_i$ are the momentum and position operators of the $i$-th electron.

The first step is to obtain the wavefunction of the electron system under the infinite quantum well potential. Considering the one-body problem for the single-particle Hamiltonian operator $\hat{H}_{\rm{mf},1}$, the Schrödinger equation is given by:
\begin{align}
  \hat{H}_{\rm{mf},1} \psi(\bm{r}) = E \psi(\bm{r}) 
  \label{1body_SHeq} \\
  \hat{H}_{\rm{mf},1} = -\frac{\hbar^2}{2m^{\ast}} \bm{\nabla}^2 + V(z)  
  \label{1body_hamiltonian} 
\end{align}
Since $V(z)$ is independent of the in-plane coordinate $\bm{\rho}$, we assume a separable form for the single-particle wavefunction $\psi(\bm{r})$:
\begin{align}
  \psi(\bm{r}) = \phi(\bm{\rho})\varphi(z)
 \label{separable_wavefunction}
\end{align}
This allows us to separate Eq.~(\ref{1body_SHeq}) into the following two eigenvalue problems:
\begin{align}
  E &= E_1 + E_2 
  \label{Etot}\\
  \left(-\frac{\hbar^2}{2m^{\ast}} \bm{\nabla}_{\parallel}^2\right) \phi(\bm{\rho}) &= E_1 \phi(\bm{\rho})
  \label{1body_SHeq_1} \\
  \left(-\frac{\hbar^2}{2m^{\ast}} \frac{d^2}{dz^2} + V(z)\right)\varphi(z) &= E_2 \varphi(z)
  \label{1body_SHeq_2} 
\end{align}
Eq.~(\ref{1body_SHeq_1}) describes a two-dimensional free electron system, with eigenenergies and eigenstates expressed in terms of the normalized area $S$ and in-plane wavevector $\bm{K}_{\parallel}$ as:
\begin{align}
  E_1 &= \frac{\hbar^2 \bm{K}_{\parallel}^2}{2m^{\ast}} 
  \label{E1} \\
  \phi(\bm{\rho}) &= \sqrt{\frac{1}{S}} e^{i\bm{K}_{\parallel}\cdot\bm{\rho}} 
  \label{phi1}
\end{align}
Eq.~(\ref{1body_SHeq_2}) corresponds to a one-dimensional infinite potential well, where the eigenenergies and eigenfunctions are given by:
\begin{align}
  E_2 &= \frac{\hbar^2 k_{zn}^2}{2m^{\ast}}  
  \label{E1} \\
  \varphi(z) 
  &= \sqrt{\frac{2}{d}}
  \left\{
  \begin{matrix}
  0 &\quad (z > d) \\
  \sin k_{zn} z &\quad (0 \leqq z \leqq d) \\
  0 &\quad (z < 0)
  \end{matrix}
  \right. \quad (n=1,2,3,\cdots) 
  \label{varphi1} 
\end{align}
where the quantized wavevector $k_{zn}$ is defined as:
\begin{align}
  k_{zn} = \frac{n\pi}{d} \quad (n= 1,2,3,\cdots)
\end{align}
Thus, the total eigenenergies and eigenstates of Eq.~(\ref{1body_SHeq}) are:
\begin{align}
  E 
  &= \frac{\hbar^2 \left(\bm{K}_{\parallel}^2 + k_{zn}^2 \right)}{2m^{\ast}} 
  \label{energy_film}\\
  \psi_{\bm{K}_{\parallel},n}(\bm{\rho},z) 
  &= \sqrt{\frac{2}{dS}} e^{i\bm{K}_{\parallel}\cdot\bm{\rho}}
  \left\{
  \begin{matrix}
  0 &\quad (z > d) \\
  \sin k_{zn} z &\quad (0 \leqq z \leqq d) \\
  0 &\quad (z < 0)
  \end{matrix}
  \right. \quad (n=1,2,3,\cdots)
  \label{eigenfunction_film}
\end{align}
Note that the wavefunctions $\psi_{\bm{K}_{\parallel},n}(\bm{\rho},z)$ form a complete orthonormal basis.

Next, we consider a plane-wave expansion of the Coulomb potential,
$f(\hat{\bm{\rho}}_i - \hat{\bm{\rho}}_j, \hat{z}_i - \hat{z}_j) = \frac{1}{\sqrt{(\hat{\bm{\rho}}_i - \hat{\bm{\rho}}_j)^2 + (\hat{z}_i - \hat{z}_j)^2}}$. As shown in Appendix~\ref{appendix_A}, this potential can be expanded in the thin-film geometry as:
\begin{align}
   f(\hat{\bm{\rho}}_i-\hat{\bm{\rho}}_j,\hat{z}_i-\hat{z}_j) 
 = \iint\frac{ d^2 \bm{k}_{\parallel} }{(2\pi)^2} \frac{1}{d}\sum_{m=-\infty}^{\infty} \frac{2\pi}{\bm{k}_{\parallel}^2 + k_{zm}^2}\left[1 - (-1)^m e^{-|\bm{k}_{\parallel}|d}\right] e^{i\bm{k}_{\parallel}\cdot(\hat{\bm{\rho}}_i - \hat{\bm{\rho}}_j)}e^{ik_{zm} (\hat{z}_i - \hat{z}_j)}
  \label{F_sub} 
\end{align}

Using Eq.~(\ref{F_sub}), Eq.~(\ref{H_mf}) can be rewritten as:
\begin{align}
  \hat{H}_{\mathrm{mf}} 
  &= \sum_{i=1}^N \frac{\hat{\bm{p}}_i^2}{2m^{\ast}} + V(z) + \frac{1}{2} \sum_{\bm{k}_{\parallel} \neq 0} \sum_{m=-\infty}^{\infty} M^2_{\bm{k}_{\parallel},m} \left(\hat{\beta}^{\dagger}_{\bm{k}_{\parallel},m} \hat{\beta}_{\bm{k}_{\parallel},m} - N\right)
\label{H_mem_basis}
\end{align}
where we define:
\begin{align}
  M_{\bm{k}_{\parallel},m} = \sqrt{\frac{e^2 \left[1 - (-1)^m e^{-|\bm{k}_{\parallel}|d} \right] }{2\varepsilon_0 V (\bm{k}_{\parallel}^2 + k_{zm}^2)}}
\end{align}
and
\begin{align}
  \hat{\beta}_{\bm{k}_{\parallel},m} &= \sum_{i=1}^{N} e^{i\bm{k}_{\parallel}\cdot\hat{\bm{\rho}}_i} e^{ik_{zm}\hat{z}_i}
    \label{beta_mem} \\
  \hat{\beta}^{\dagger}_{\bm{k}_{\parallel},m} &= \sum_{i=1}^{N} e^{-i\bm{k}_{\parallel}\cdot\hat{\bm{\rho}}_i} e^{-ik_{zm}\hat{z}_i}
    \label{beta_dagger_mem} 
\end{align}
The $\bm{k}_{\parallel} = 0$ term is excluded, as it cancels due to interaction with the uniform positive background charge, consistent with the jellium model~\cite{fetter2012quantum}.

Next, as in the bulk case, we introduce the Bohm-Pines Hamiltonian with the following  degrees of freedom based on the Bohm–Pines theory~\cite{bohm1951collective1,pines1952collective2,bohm1953collective3,pines1953collective4}:
\begin{align}
  \hat{H}_{\mathrm{mf}} 
  &= \sum_{i=1}^N \frac{\hat{\bm{p}}_i^2}{2m^{\ast}}+ V(z) +  \frac{1}{2} \sum_{|\bm{k}_{\parallel}| > k_c} \sum_{m=-\infty}^{\infty} {M}^2_{\bm{k}_{\parallel},m} \lr{\hat{\beta}^{\dagger}_{\bm{k}_{\parallel},m}\hat{\beta}_{\bm{k}_{\parallel},m} - N}\nonumber \\
  &+ \frac{1}{2} \sum_{0 < |\bm{k}_{\parallel}| \leq k_c} \sum_{m=-\infty}^{\infty} {M}^2_{\bm{k}_{\parallel},m} \lr{\hat{\beta}^{\dagger}_{\bm{k}_{\parallel},m}\hat{\beta}_{\bm{k}_{\parallel},m} - N}  \nonumber \\
  \rightarrow \hat{H'}_{\mathrm{mf}}
  &= \sum_{i=1}^N \frac{\hat{\bm{p}}_i^2}{2m^{\ast}}+ V(z) + \frac{1}{2} \sum_{|\bm{k}_{\parallel}| > k_c} \sum_{m=-\infty}^{\infty} {M}^2_{\bm{k}_{\parallel},m} \lr{\hat{\beta}^{\dagger}_{\bm{k}_{\parallel},m}\hat{\beta}_{\bm{k}_{\parallel},m} - N}\nonumber \\
  &+ \frac{1}{2} \sum_{|\bm{k}_{\parallel}| \leq k_c} \sum_{m=-\infty}^{\infty} \blr{\lr{{M}_{\bm{k}_{\parallel},m}\hat{\beta}^{\dagger}_{\bm{k}_{\parallel},m}-\hat{\mathcal{P}}^{\dagger}_{\bm{k}_{\parallel},m}}\lr{{M}_{\bm{k}_{\parallel},m}\hat{\beta}_{\bm{k}_{\parallel},m}-\hat{\mathcal{P}}_{\bm{k}_{\parallel},m}} - {M}^2_{\bm{k}_{\parallel},m} N} \nonumber \\
  &= \sum_{i=1}^N \frac{\hat{\bm{p}}_i^2}{2m^{\ast}}+ V(z) + \frac{1}{2} \sum_{|\bm{k}_{\parallel}| > k_c} \sum_{m=-\infty}^{\infty} {M}^2_{\bm{k}_{\parallel},m} \lr{\hat{\beta}^{\dagger}_{\bm{k}_{\parallel},m}\hat{\beta}_{\bm{k}_{\parallel},m} - N}\nonumber \\
  &+ \frac{1}{2} \sum_{\mu} \blr{\lr{{M}_{\mu}\hat{\beta}^{\dagger}_{\mu}-\hat{\mathcal{P}}^{\dagger}_{\mu}}\lr{{M}_{\mu}\hat{\beta}_{\mu}-\hat{\mathcal{P}}_{\mu}} - {M}^2_{\mu} N}
\label{H_mem_basis_3}
\end{align}
The $k_c$ is the cutoff wavenumber, uniquely determined to minimize the ground state energy.
For brevity, we have introduced the index $\mu \clr{=\bm{k}_{\parallel} \lr{|\bm{k}_{\parallel}|<k_c},m}$ to label the states.\\
We assume that $\hat{\mathcal{P}}^{\dagger}_{\mu} = \hat{\mathcal{P}}_{-\mu},\hat{\mathcal{Q}}^{\dagger}_{\mu} = \hat{\mathcal{Q}}_{-\mu}$ and that they satisfy the following canonical commutation relation:
\begin{align}
  \blr{\hat{\mathcal{Q}}_{\mu},\hat{\mathcal{P}}_{\mu'}} = i\hbar \delta_{\mu\mu'}
\end{align}
The two Hamiltonians $\hat{H}_{\mathrm{mf}}$ and $\hat{H'}_{\mathrm{mf}}$ describe the same physical system under the following subsidiary condition on the many-body wavefunction $\Psi$:
\begin{align}
  \hat{\mathcal{P}}^{\dagger}_{\mu} \Psi = 0 
\end{align}

Furthermore, performing a unitary transformation using the operator
\begin{align}
  \hat{U} &= e^{\lr{\frac{i}{\hbar}}\hat{S}} 
  \label{mem_unitary} \\
  \hat{S} &= \sum_{\mu} {M}_{\mu} \hat{\mathcal{Q}}_{\mu}\hat{\beta}_{\mu} 
\end{align}
each operator transforms as follows:
\begin{align}
  \lr{\hat{\mathcal{P}}_{\mu}}_U &= \hat{\mathcal{P}}_{\mu} + M_{\mu}\hat{\beta}_{\mu}
  \label{Unitary_trans_pl_momentum}\\
  \lr{\hat{\bm{p}}_i}_U &= \hat{\bm{p}}_i + i\sum_{\mu} M_{\mu}\hat{\mathcal{Q}}_{\mu} 
  \begin{pmatrix}
    \bm{k}_{\parallel} \\
    k_{zm}
  \end{pmatrix}
  e^{i\bm{k}_{\parallel}\cdot\hat{\bm{\rho}}_{i}} e^{ik_{zm} \hat{z}_i}
  \label{Unitary_trans_ele_momentum}
\end{align}

From this, the first term in Eq.~(\ref{H_mem_basis_3}) transforms as:
\begin{align}
\lr{\sum_{i=1}^N \frac{\hat{\bm{p}}_i^2}{2m^{\ast}}}_U 
  =& \sum_{i=1}^N \frac{\hat{\bm{p}}_i^2}{2m^{\ast}} + \frac{i}{2m^{\ast}}\sum_{i=1}^N\sum_{\mu} M_{\mu}\hat{\mathcal{Q}}_{\mu}
  \begin{pmatrix}
    \bm{k}_{\parallel} \\
    k_{zm}
  \end{pmatrix}
  \cdot
  \clr{
    \hat{\bm{p}}_{i}
  ,
  e^{i\bm{k}_{\parallel}\cdot\hat{\bm{\rho}}_{i}} e^{ik_{zm} \hat{z}_i}
  } 
  \nonumber \\
 &+ \frac{1}{2m^{\ast}}\sum_{i=1}^N \sum_{\mu\mu'}M_{\mu}M_{\mu'}\hat{\mathcal{Q}}_{\mu}\hat{\mathcal{Q}}_{\mu'}
 \blr{-\lr{\bm{k}_{\parallel}\cdot\bm{k'}_{\parallel}} -k_{zm} k_{zm'} }e^{i\lr{\bm{k}_{\parallel}+\bm{k'}_{\parallel}}\cdot\hat{\bm{\rho}}_{i}} e^{i\lr{k_{zm}+k_{zm'}}\hat{z}_i}
\label{1st term_unitary}
\end{align}
By redefining the final term in Eq.~(\ref{1st term_unitary}) with $\mu' = -\mu$ and applying the random phase approximation (RPA), only the terms with $\mu = \mu'$ remain, yielding:
\begin{align}
 & \frac{1}{2m^{\ast}}\sum_{i=1}^N \sum_{\mu\mu'} M_{\mu}M'_{-\mu'}\hat{\mathcal{Q}}_{\mu}\hat{\mathcal{Q}}_{-\mu'}
 \blr{\lr{\bm{k}_{\parallel}\cdot\bm{k'}_{\parallel}} + k_{zm} k_{zm'} }e^{i\lr{\bm{k}_{\parallel}-\bm{k'}_{\parallel}}\cdot\hat{\bm{\rho}}_{i}} e^{i\lr{k_{zm}-k_{zm'}}\hat{z}_i}
 \nonumber \\
\cong & \frac{N}{2m^{\ast}}\sum_{\mu} M_{\mu}M'_{-\mu}\hat{\mathcal{Q}}_{\mu}\hat{\mathcal{Q}}_{-\mu}
\blr{|\bm{k}_{\parallel}|^2 + k^2_{zm} } \ \ \ \ \ (\mathrm{RPA})
\nonumber \\
 =    & \sum_{\mu} \frac{Ne^2}{4m^{\ast}\varepsilon_0 V}\clr{1-\lr{-1}^m e^{-|\bm{k}_{\parallel}|d}}\hat{\mathcal{Q}}^{\dagger}_{\mu} \hat{\mathcal{Q}}_{\mu}
 \nonumber \\
 =    & \frac{1}{2} \sum_{\mu} \omega_{\mu}^{(\rm{pl}) 2} \hat{\mathcal{Q}}^{\dagger}_{\mu} \hat{\mathcal{Q}}_{\mu}
 \label{1st term_unitary_RPA}
\end{align}
where the plasmon dispersion $\omega_{\mu}^{(\rm{pl})}$ is given by
\begin{align}
  \omega_{\mu}^{(\rm{pl})} = \frac{\omega_{\rm{pl}}}{\sqrt{2}}\sqrt{1-\lr{-1}^m e^{-|\bm{k}_{\parallel}|d}} 
\end{align}
which coincides with the electrostatic thin-film plasmon dispersion relation given by Ferrell~\cite{ferrell1958predicted}.\\

The fourth term in Eq.~(\ref{H_mem_basis_3}) becomes:
\begin{align}
  &\frac{1}{2} \sum_{\mu} \blr{\lr{{M}_{\mu}\hat{\beta}^{\dagger}_{\mu}-\lr{\hat{\mathcal{P}}^{\dagger}_{\mu}}_U}\lr{{M}_{\mu}\hat{\beta}_{\mu}-\lr{\hat{\mathcal{P}}_{\mu}}_U} - {M}^2_{\mu} N}
  \nonumber \\
 =&\frac{1}{2} \sum_{\mu} \blr{\hat{\mathcal{P}}^{\dagger}_{\mu} \hat{\mathcal{P}}_{\mu}- {M}^2_{\mu} N}
\end{align}

Summarizing the above results, Bohm–Pines Hamiltonian in a metallic Film is expressed as follows:
\begin{align}
  \hat{H'}_{\mathrm{mf}}       &= \hat{H}_{\mathrm{mf},\rm{pl}} + \hat{H}_{\mathrm{mf},\rm{e}} +\hat{H}_{\mathrm{mf},\rm{e-pl}} + \hat{H}_{\mathrm{mf},C'} 
  \label{mem_hamiltonian_final} \\
  \hat{H}_{\mathrm{mf},\rm{pl}}  &= \sum_{\mu}  \frac{1}{2} \clr{ \hat{\mathcal{P}}^{\dagger}_{\mu}\hat{\mathcal{P}}_{\mu}+ \omega_{\mu}^{(\rm{pl}) 2} \hat{\mathcal{Q}}^{\dagger}_{\mu}\hat{\mathcal{Q}}_{\mu}} 
  \label{mem_hamiltonian_pl} \\
  \hat{H}_{\mathrm{mf},\rm{e}}   &= \sum_{i=1}^N \frac{\hat{\bm{p}}_i^2}{2m^{\ast}} +V(z) -\sum_{\mu} \frac{N}{2} {M}^2_{\mu} 
  \label{mem_hamiltonian_e} \\
  \hat{H}_{\mathrm{mf},\rm{e-pl}}&=  \frac{i}{2m^{\ast}}\sum_{i=1}^N \sum_{\mu} M_{\mu}\hat{\mathcal{Q}}_{\mu}
                                    \begin{pmatrix}
                                      \bm{k}_{\parallel} \\
                                      k_{zm}
                                    \end{pmatrix}
                                    \cdot
                                    \clr{
                                      \hat{\bm{p}}_{i}
                                    ,
                                    e^{i\bm{k}_{\parallel}\cdot\hat{\bm{\rho}}_{i}} e^{ik_{zm} \hat{z}_i}
                                    }  
  \label{mem_hamiltonian_e-pl} \\
  \hat{H}_{\mathrm{mf},C'}       &=   \sum_{|\bm{k}_{\parallel}| > k_c} \sum_{m=-\infty}^{\infty} \frac{1}{2}M_{\bm{k}_{\parallel},m}^{2} \lr{\hat{\beta}^{\dagger}_{\bm{k}_{\parallel},m}\hat{\beta}_{\bm{k}_{\parallel},m}-N} 
  \label{mem_hamiltonian_C'}
\end{align}

\subsection{ Second-Quantization of Bohm–Pines Hamiltonian in a metallic film }
Let us now consider Eq.~(\ref{mem_hamiltonian_pl}).  
Similar to bulk plasmons, thin-film plasmons also take the form of independent harmonic oscillators and are expected to admit second quantization.  
By defining the operators $\hat{\mathcal{A}}^{\dagger}_{\mu},\hat{\mathcal{A}}_{\mu}$ as
\begin{align}
  \hat{\mathcal{A}}^{\dagger}_{\mu} &= \sqrt{\frac{\omega^{(\rm{pl})}_{\mu}}{2\hbar}}\hat{\mathcal{Q}}_{\mu} -i \sqrt{\frac{1}{2\hbar\omega^{(\rm{pl})}_{\mu}}} \hat{\mathcal{P}}_{\mu} 
  \label{a_k*_bulk} \\
  \hat{\mathcal{A}}_{\mu}           &= \sqrt{\frac{\omega^{(\rm{pl})}_{\mu}}{2\hbar}}\hat{\mathcal{Q}}_{\mu} +i \sqrt{\frac{1}{2\hbar\omega^{(\rm{pl})}_{\mu}}} \hat{\mathcal{P}}_{\mu}
  \label{a_k_bulk}
\end{align}
the operators $\hat{\mathcal{A}}^{\dagger}_{\mu},\hat{\mathcal{A}}_{\mu}$ satisfy the following commutation relation:
\begin{align}
  \blr{\hat{\mathcal{A}}_{\mu} ,\hat{\mathcal{A}}^{\dagger}_{\mu}}
   &= \hat{\mathcal{A}}_{\mu}\hat{\mathcal{A}}^{\dagger}_{\mu} - \hat{\mathcal{A}}^{\dagger}_{\mu}\hat{\mathcal{A}}_{\mu} \nonumber \\
   &= -\frac{i}{2\hbar}\hat{\mathcal{Q}}_{\mu}\hat{\mathcal{P}}_{\mu}+\frac{i}{2\hbar}\hat{\mathcal{P}}_{\mu}\hat{\mathcal{Q}}_{\mu}-\frac{i}{2\hbar}\hat{\mathcal{Q}}_{\mu}\hat{\mathcal{P}}_{\mu} +\frac{i}{2\hbar}\hat{\mathcal{P}}_{\mu}\hat{\mathcal{Q}}_{\mu} \nonumber \\
   &= \frac{i}{\hbar}\blr{\hat{\mathcal{P}}_{\mu},\hat{\mathcal{Q}}_{\mu}} \nonumber \\
   &= 1 \label{bosonic_mem}
\end{align}
and thus obey the bosonic commutation relation.  
Therefore, the Hamiltonian $\hat{H}_{\mathrm{mem},\rm{pl}}$ in Eq.~(\ref{mem_hamiltonian_pl}) can be rewritten as
\begin{align}
  &\hat{H}_{\mathrm{mem},\rm{pl}} \nonumber \\
  =&\sum_{\mu} \frac{1}{2} \clr{ \hat{\mathcal{P}}^{\dagger}_{\mu}\hat{\mathcal{P}}_{\mu}+ \omega^{(\rm{pl}) 2}_{\mu} \hat{\mathcal{Q}}^{\dagger}_{\mu}\hat{\mathcal{Q}}_{\mu}} \nonumber \\
  =&\sum_{\mu} \hbar \omega^{(\rm{pl})}_{\mu} \lr{\hat{\mathcal{A}}^{\dagger}_{\mu}\hat{\mathcal{A}}_{\mu}+\frac{1}{2}}
\end{align}
which yields a second-quantized representation of thin-film plasmons.

The Hamiltonian $\hat{H}_{\mathrm{mf},\rm{e}}$ in Eq.~(\ref{mem_hamiltonian_e}) describes the motion of free fermionic particles confined in a thin film (ignoring constant terms).  
Following a similar argument as in the bulk case, it is shown that bosonic excitations, namely ``electron–hole pair excitations,'' can emerge from the above fermionic motion.

First, the Hamiltonian can be expressed using the field operators $\hat{\Psi}^{\dagger}(\bm{x}),\hat{\Psi}(\bm{x})$ as
\begin{align}
  \hat{H}_{\mathrm{mf},\rm{e}} = \int d\bm{x} \hat{\Psi}^{\dagger}(\bm{x})\lr{-\frac{\hbar^2 \bm{\nabla}_{\bm{x}}^2}{2m^{\ast}} + V(z)}\hat{\Psi}(\bm{x})
\end{align}
The field operators $\hat{\Psi}^{\dagger}(\bm{x}),\hat{\Psi}(\bm{x})$ are defined in terms of the wave functions $\psi_{\bm{K}_{\parallel},n}(\bm{x})$ introduced in Eq.~\ref{eigenfunction_film} and the creation and annihilation operators $\hat{c}^{\dagger}_{\bm{K}_{\parallel},n},\hat{c}_{\bm{K}_{\parallel},n}$ as
\begin{align}
  \hat{\Psi}          (\bm{x}) &= \sum_{\bm{K}_{\parallel},n}\psi_{\bm{K}_{\parallel},n}(\bm{x}) \hat{c}_{\bm{K}_{\parallel},n}                    
  \\
  \hat{\Psi}^{\dagger}(\bm{x}) &= \sum_{\bm{K}_{\parallel},n}\psi^{\ast}_{\bm{K}_{\parallel},n}(\bm{x}) \hat{c}^{\dagger}_{\bm{K}_{\parallel},n}
\end{align}
Accordingly, the Hamiltonian is second-quantized as
\begin{align}
  \hat{H}_{\mathrm{mf},\rm{e}} 
  &= \sum_{\bm{K}_{\parallel},\bm{K'}_{\parallel}}\sum_{n,n'} \clr{\int d\bm{x} \psi_{\bm{K'}_{\parallel},n'}^{*}(\bm{x})\lr{-\frac{\hbar^2 \bm{\nabla}_{\bm{x}}^2}{2m}+ V(z)}\psi_{\bm{K}_{\parallel},n}(\bm{x})}\hat{c}^{\dagger}_{\bm{K'}_{\parallel},n'}\hat{c}_{\bm{K}_{\parallel},n} \nonumber \\
  &= \sum_{\bm{K}_{\parallel},n} \frac{\hbar^2 \lr{\bm{K}_{\parallel}^2+k_{z,n}^2 }}{2m^{\ast}} \hat{c}^{\dagger}_{\bm{K}_{\parallel},n}\hat{c}_{\bm{K}_{\parallel},n} \nonumber \\
  &= \sum_{\bm{K}_{\parallel},n} \varepsilon_{\bm{K}_{\parallel},n} \hat{c}^{\dagger}_{\bm{K}_{\parallel},n}\hat{c}_{\bm{K}_{\parallel},n}
\end{align}
By Pauli's exclusion principle, only one particle can occupy a given state.  
The energy of the Fermi-degenerate state is given by
\begin{align}
  E_{\mathrm{F}} = \sum_{|\bm{K}_{\parallel}|\leq k_{\mathrm{F},n}}\varepsilon_{\bm{K}_{\parallel},n}
\end{align}
Here, $k_{\mathrm{F},n}$ denotes the radius of the Fermi circle in the $n$th subband (i.e., the Fermi wave number in that subband), and is given by
\begin{align}
  k_{\mathrm{F},n} = k_{\mathrm{F}} -\frac{n\pi}{d}
  \label{Fermi-wavenumber_n}
\end{align}
Thus, the Hamiltonian can be rewritten as
\begin{align}
  \hat{H}_{\mathrm{mf},\rm{e}}  
  &= \sum_{ \bm{K}_{\parallel},n} \varepsilon_{\bm{K}_{\parallel},n} \hat{c}^{\dagger}_{\bm{K}_{\parallel},n}\hat{c}_{\bm{K}_{\parallel},n} -\sum_{|\bm{K}_{\parallel}|\leq k_{\mathrm{F},n}}\varepsilon_{\bm{K}_{\parallel},n} + E_{\mathrm{F}} \nonumber \\
  &= \sum_{|\bm{K}_{\parallel}|\leq k_{\mathrm{F},n}} \varepsilon_{\bm{K}_{\parallel},n} \lr{\hat{c}^{\dagger}_{\bm{K}_{\parallel},n}\hat{c}_{\bm{K}_{\parallel},n}-1} + \sum_{|\bm{K}_{\parallel}| > k_{\mathrm{F},n}}\varepsilon_{\bm{K}_{\parallel},n}\hat{c}^{\dagger}_{\bm{K}_{\parallel},n}\hat{c}_{\bm{K}_{\parallel},n} + E_{\mathrm{F}} \nonumber \\
  &= \sum_{|\bm{K}_{\parallel}|\leq k_{\mathrm{F},n}} \lr{-\varepsilon_{\bm{K}_{\parallel},n}}\hat{c}_{\bm{K}_{\parallel},n}\hat{c}^{\dagger}_{\bm{K}_{\parallel},n}   + \sum_{|\bm{K}_{\parallel}| > k_{\mathrm{F},n}}\varepsilon_{\bm{K}_{\parallel},n}\hat{c}^{\dagger}_{\bm{K}_{\parallel},n}\hat{c}_{\bm{K}_{\parallel},n} + E_{\mathrm{F}}    
\end{align}

Now, we redefine the electron and hole creation and annihilation operators as follows:
\begin{align}
  \hat{h}_{\nu_{\rm{h}}}= \hat{c}^{\dagger}_{\nu_{\rm{h}}}, \hat{h}^{\dagger}_{\nu_{\rm{h}}}= \hat{c}_{\nu_{\rm{h}}} &
  \ \ \ \clr{\nu_{\rm{h}} = \bm{K }_{\parallel},n \lr{|\bm{K }_{\parallel}|\leq k_{\mathrm{F},n }}} 
  \label{h_i}\\
  \hat{e}_{\nu_{\rm{e}}}= \hat{c}_{\nu_{\rm{e}}}, \hat{e}^{\dagger}_{\nu_{\rm{e}}}= \hat{c}^{\dagger}_{\nu_{\rm{e}}} &
  \ \ \ \clr{\nu_{\rm{e}} = \bm{K'}_{\parallel},n' \lr{|\bm{K'}_{\parallel}| > k_{\mathrm{F},n' }}}
  \label{e_i}
\end{align}
These redefined operators $\hat{h}^{\dagger},\hat{h},\hat{e}^{\dagger},\hat{e}$ satisfy the fermionic anticommutation relations.

Using these operators, the Hamiltonian $\hat{H}_{\mathrm{mf},\rm{e}}$ becomes
\begin{align}
  \hat{H}_{\mathrm{mf},\rm{e}} =\sum_{\nu_{\rm{h}}} \lr{-\varepsilon_{\nu_{\rm{h}}}}\hat{h}^{\dagger}_{\nu_{\rm{h}}}\hat{h}_{\nu_{\rm{h}}} +\sum_{\nu_{\rm{e}}} \varepsilon_{\nu_{\rm{e}}}\hat{e}^{\dagger}_{\nu_{\rm{e}}}\hat{e}_{\nu_{\rm{e}}} + E_{\mathrm{F}} 
\label{electron-energy-2}
\end{align}
That is, the total energy of individual excitations is determined by counting the number of holes inside the Fermi circle and the number of electrons outside it for each subband.  
If we define the energy relative to the Fermi energy $\varepsilon_{\mathrm{F}}$ as $\tilde{\varepsilon}_{\bm{K}_{\parallel},n}$, then
\begin{align}
  \varepsilon_{\bm{K}_{\parallel},n} = \varepsilon_{\mathrm{F}} +  \tilde{\varepsilon}_{\bm{K}_{\parallel},n}
\label{fermmi-2}
\end{align}
Substituting Eq.~(\ref{fermmi-2}) into Eq.~(\ref{electron-energy-2}), we get
\begin{align}
  \hat{H}_{\mathrm{mf},\rm{e}}&=\sum_{\nu_{\rm{h}}} \lr{-\tilde{\varepsilon}_{\nu_{\rm{h}}}}
                                   \hat{h}^{\dagger}_{\nu_{\rm{h}}}\hat{h}_{\nu_{\rm{h}}} 
                               +\sum_{\nu_{\rm{e}}} \tilde{\varepsilon}_{\nu_{\rm{e}}}
                                   \hat{e}^{\dagger}_{\nu_{\rm{e}}}\hat{e}_{\nu_{\rm{e}}} \nonumber \\ 
                              &+\varepsilon_{\mathrm{F}}\clr{-\sum_{\nu_{\rm{h}}}\hat{h}^{\dagger}_{\nu_{\rm{h}}}\hat{h}_{\nu_{\rm{h}}} 
                               +\sum_{\nu_{\rm{e}}} \hat{e}^{\dagger}_{\nu_{\rm{e}}}\hat{e}_{\nu_{\rm{e}}} } + E_{\mathrm{F}} \nonumber \\
                                  &=\sum_{\nu_{\rm{h}}} \lr{-\tilde{\varepsilon}_{\nu_{\rm{h}}}}\hat{h}^{\dagger}_{\nu_{\rm{h}}}\hat{h}_{\nu_{\rm{h}}} +\sum_{\nu_{\rm{e}}} \tilde{\varepsilon}_{\nu_{\rm{e}}}\hat{e}^{\dagger}_{\nu_{\rm{e}}}\hat{e}_{\nu_{\rm{e}}} + E_{\mathrm{F}} \ \ \ \lr{\text{since particle number is conserved}} \nonumber \\
                                  &=\sum_{\nu_{\rm{h}}} \varepsilon_{\nu_{\rm{h}}}^{(h)}\hat{h}^{\dagger}_{\nu_{\rm{h}}}\hat{h}_{\nu_{\rm{h}}}         +\sum_{\nu_{\rm{h}}} \varepsilon_{\nu_{\rm{e}}}^{(e)}\hat{e}^{\dagger}_{\nu_{\rm{e}}}\hat{e}_{\nu_{\rm{h}}} + E_{\mathrm{F}}  \\
                                  &\ \ \ \lr{\varepsilon_{\nu_{\rm{h}}}^{(h)}=\varepsilon_{\mathrm{F}}-\varepsilon_{\bm{K}_{\parallel},n},\varepsilon_{\nu_{\rm{e}}}^{(e)}=\varepsilon_{\bm{K'}_{\parallel},n'}-\varepsilon_{\mathrm{F}}} \nonumber
\label{electron-energy-3}
\end{align}
Next, we define the electron–hole pair's creation and annihilation operators as
\begin{align}
  \hat{\mathcal{B}}^{\dagger}_{\nu_{\rm{h}},\nu_{\rm{e}}} =\hat{h}^{\dagger}_{\nu_{\rm{h}}}\hat{e}^{\dagger}_{\nu_{\rm{e}}}, \ \ \hat{\mathcal{B}}_{\nu_{\rm{h}},\nu_{\rm{e}}} =\hat{h}_{\nu_{\rm{h}}} \hat{e}_{\nu_{\rm{e}}}
\end{align}
Their commutation relation is given by
\begin{align}
  &\blr{\hat{\mathcal{B}}_{{\nu}_{\rm{h}},{\nu}_{\rm{e}}}, \hat{\mathcal{B}}^{\dagger}_{{\nu'}_{\rm{h}},{\nu'}_{\rm{e}}}} \nonumber \\
 =& \delta_{{\nu}_{\rm{h}},{\nu'}_{\rm{h}}}\delta_{{\nu}_{\rm{e}},{\nu'}_{\rm{e}}} -\delta_{{\nu}_{\rm{h}},{\nu'}_{\rm{h}}} \hat{e}^{\dagger}_{{\nu'}_{\rm{e}}}\hat{e}_{{\nu}_{\rm{e}}}-\delta_{{\nu}_{\rm{e}},{\nu'}_{\rm{e}}}\hat{h}^{\dagger}_{{\nu'}_{\rm{h}}}\hat{h}_{{\nu}_{\rm{h}}}
\end{align}
and
\begin{align}
  & \sum_{{\nu}_{\rm{h}}} \sum_{{\nu}_{\rm{e}}} \lr{\varepsilon_{{\nu}_{\rm{h}}}^{(h)} + \varepsilon_{{\nu}_{\rm{e}}}^{(e)}} \hat{\mathcal{B}}^{\dagger}_{{\nu}_{\rm{h}},{\nu}_{\rm{e}}}\hat{\mathcal{B}}_{{\nu}_{\rm{h}},{\nu}_{\rm{e}}} \nonumber \\
  &=\sum_{{\nu}_{\rm{h}}} \varepsilon_{{\nu}_{\rm{h}}}^{(h)}
    \lr{\sum_{{\nu}_{\rm{e}}}\hat{e}^{\dagger}_{{\nu}_{\rm{e}}}\hat{e}_{{\nu}_{\rm{e}}}}
    \hat{h}^{\dagger}_{{\nu}_{\rm{h}}}\hat{h}_{{\nu}_{\rm{h}}}    
    +\sum_{{\nu}_{\rm{e}}} \varepsilon_{{\nu}_{\rm{e}}}^{(e)} 
\lr{\sum_{{\nu}_{\rm{h}}}\hat{h}^{\dagger}_{{\nu}_{\rm{h}}}\hat{h}_{{\nu}_{\rm{h}}}}\hat{e}^{\dagger}_{{\nu}_{\rm{e}}}\hat{e}_{{\nu}_{\rm{e}}} \nonumber \\
  &=\hat{N}_{\mathrm{e-h}}\lr{\sum_{{\nu}_{\rm{h}}} \varepsilon_{{\nu}_{\rm{h}}}^{(h)}\hat{h}^{\dagger}_{{\nu}_{\rm{h}}}\hat{h}_{{\nu}_{\rm{h}}} +\sum_{{\nu}_{\rm{e}}}\varepsilon_{{\nu}_{\rm{e}}}^{(e)} \hat{e}^{\dagger}_{{\nu}_{\rm{e}}}\hat{e}_{{\nu}_{\rm{e}}}  } 
\label{relation-1}
\end{align}
Here, $\hat{N}_{\mathrm{e-h}}$ is the number operator for electron–hole pair excitations.  
Although one could consider states with multiple excitations, we restrict ourselves here to the single electron–hole pair excitation case $(\hat{N}_{\mathrm{e-h}}=1)$.  
Since the sum of in-plane momentum vectors vanishes in the ground state, the in-plane momentum $\bm{k}_{\parallel}$ of an electron–hole pair is given by
\begin{align}
  \bm{k}_{\parallel} = \bm{K'}_{\parallel}-\bm{K}_{\parallel} 
  \label{kK'K}
\end{align}
Because the quantum state of an electron–hole pair is specified by the index $\nu \clr{= \nu_{\rm{h}},\nu_{\rm{e}}}$, from Eq.~(\ref{relation-1}) we obtain
\begin{align}
  &\hat{H}_{\mathrm{mf},\rm{e}} \nonumber \\
 =& \sum_{{\nu}_{\rm{h}}} \varepsilon_{{\nu}_{\rm{h}}}^{(h)}\hat{h}^{\dagger}_{{\nu}_{\rm{h}}}\hat{h}_{{\nu}_{\rm{h}}} +\sum_{{\nu}_{\rm{e}}}\varepsilon_{{\nu}_{\rm{e}}}^{(e)} \hat{e}^{\dagger}_{{\nu}_{\rm{e}}}\hat{e}_{{\nu}_{\rm{e}}} +E_{\mathrm{F}} \nonumber \\
 =& \sum_{\nu} \frac{\hbar^2}{2m^{\ast}}\blr{\lr{\bm{k}_{\parallel}^2+2\bm{K}_{\parallel}\cdot\bm{k}_{\parallel}}+\lr{\frac{\pi}{d}}^2\lr{{n'}^2-n^2} }\hat{\mathcal{B}}^{\dagger}_{\nu}\hat{\mathcal{B}}_{\nu} + E_{\mathrm{F}} \nonumber \\
 =& \sum_{\nu} \hbar\omega^{(\rm{e-h})}_{\nu} \hat{\mathcal{B}}^{\dagger}_{\nu}\hat{\mathcal{B}}_{\nu} + E_{\mathrm{F}} 
\label{electron-energy-4}
\end{align}
Only the states that satisfy
\begin{align}
  |\bm{K}_{\parallel}|                    &\leq k_{\mathrm{F},n } \nonumber \\
  |\bm{K}_{\parallel}+\bm{k}_{\parallel}| &   > k_{\mathrm{F},n'} \nonumber 
\end{align}
are counted as valid $\nu$-states, and the excitation energy $\hbar\omega^{(\rm{e-h})}_{\nu}$ of an electron–hole pair is expressed as
\begin{align}
  \hbar\omega^{(\rm{e-h})}_{\nu} = \frac{\hbar^2}{2m^{\ast}}\lr{\bm{k}_{\parallel}^2+2\bm{K}_{\parallel}\cdot\bm{k}_{\parallel}}+\frac{\hbar^2}{2m^{\ast}}\lr{\frac{\pi}{d}}^2\lr{{n'}^2-n^2} 
\end{align}

Although the total current density operator of the system $\hat{\bm{J}}$ is generally defined as
\begin{align}
    \hat{\bm{J}}(\bm{r}) = \sum_{i=1}^{N}\blr{\frac{e\clr{\hat{\bm{p}}_i\delta(\bm{r}-\hat{\bm{r}}_i)+\delta(\bm{r}-\hat{\bm{r}}_i)\hat{\bm{p}}_i}}{2m^{\ast}}} \nonumber 
\end{align}
in the derivation of the Bohm–Pines Hamiltonian, the electron momentum operator is transformed via a unitary operator, and thus the total current density operator $\hat{\bm{J}}$ must be rewritten accordingly.

As a result, the total current density operator $\hat{\bm{J}}$ is separated into a plasmonic component $\hat{\bm{J}}_{\rm{p}}$ and an individual excitation component $\hat{\bm{J}}_{\rm{e}}$ as follows:
\begin{align}
  \hat{\bm{J}}(\bm{r}) &= \hat{\bm{J}}_{\rm{p}}(\bm{r}) + \hat{\bm{J}}_{\rm{{e}}}(\bm{r})
  \label{totalJ} \\
  \hat{\bm{J}}_{\rm{{p}}}(\bm{r}) &= \frac{Ne^2 i}{m^{\ast} V } \sqrt{\frac{1}{\varepsilon_0 V}}
  \sum_{\mu}
  \frac{\omega_{\mu}^{(\rm{pl})}}{\omega_{\rm{pl}}} \hat{\mathcal{Q}}_{\mu}
  \bm{\varepsilon}_{\bm{k}_{\parallel},m}e^{i\bm{k}_{\parallel}\cdot\bm{\rho}} e^{ik_{zm} z}
  \label{Jp}     \\
  \hat{\bm{J}}_{\rm{e}}(\bm{r})   &= \sum_{i=1}^{N} \blr{\frac{e\clr{\hat{\bm{p}}_i\delta(\bm{r}-\hat{\bm{r}}_i)+\delta(\bm{r}-\hat{\bm{r}}_i)\hat{\bm{p}}_i}}{2m^{\ast}}}
  \label{Je}     
\end{align}
where $\bm{\varepsilon}_{\bm{k}_{\parallel},m}$ means the unit vector of $\bm{k} = \begin{pmatrix}
   \bm{k}_{\parallel} \\ 
   k_{zm}
\end{pmatrix}$. The plasmon-induced current density operator $\hat{\bm{J}}_{\rm{p}}$ in Eq.~(\ref{Jp}) becomes
\begin{align}
  \hat{\bm{J}}_{\rm{{p}}}(\bm{r}) 
  &= 
  \sum_{\mu} 
  \lr{\bm{\mathcal{J}}_{p;\mu} (\bm{r}) \hat{\mathcal{A}}_{\mu} + \rm{H.c.}} \nonumber 
\end{align}
where
\begin{align}
  \bm{\mathcal{J}}_{p;\mu} (\bm{r}) &= i \varepsilon_0  \omega_{\rm{pl}} \sqrt{ \frac{\hbar \omega_{\mu}^{(\rm{pl})}}{2\varepsilon_0 V}}
  \bm{\varepsilon}_{\bm{k}_{\parallel},m}e^{i\bm{k}_{\parallel}\cdot\bm{\rho}} e^{ik_{zm} z}
\end{align}

and individual-excitation-induced current density operator $\hat{\bm{J}}_{\rm{e}}$ in Eq.~(\ref{Je}) becomes 
\begin{align}
  \hat{\bm{J}}_{\rm{e}} (\bm{r})
 =& \sum_{\nu}\lr{\bm{\mathcal{J}}_{e;\nu}(\bm{r})\hat{\mathcal{B}}_{\nu} + \rm{H.c.}} 
   \label{J_e_2ndquantization}
\end{align}
where 
\begin{align}
  \bm{\mathcal{J}}_{e;\nu}(\bm{r}) &= 
  \frac{\hbar e}{m^{\ast} V} 
   \blr{
   \begin{matrix}
       \lr{2\bm{K}_{\parallel} + \bm{k}_{\parallel}} \sin k_{z,n'}z \sin k_{z,n}z \\
        \frac{k_{zn'}-k_{zn}}{2i}\sin\lr{k_{zn'}+k_{zn}}z -\frac{k_{zn'}+k_{zn}}{2i}\sin\lr{k_{zn'}-k_{zn}}z  
   \end{matrix}
    } e^{i\bm{k}_{\parallel}\cdot\bm{\rho}} 
\end{align}

The term $\hat{H}_{\mathrm{mf},\rm{e-pl}}$ in Eq.~(\ref{mem_hamiltonian_e-pl}) represents the residual longitudinal interaction between plasmons and individual excitations. It can be written as
\begin{align}
      \hat{H}_{\mathrm{mf},\rm{e-pl}}&=  \frac{i}{2m^{\ast}}\sum_{i=1}^N \sum_{\mu} M_{\mu}\hat{\mathcal{Q}}_{\mu}
                                    \begin{pmatrix}
                                      \bm{k}_{\parallel} \\
                                      k_{zm}
                                    \end{pmatrix}
                                    \cdot
                                    \clr{
                                      \hat{\bm{p}}_{i}
                                    ,
                                    e^{i\bm{k}_{\parallel}\cdot\hat{\bm{\rho}}_{i}} e^{ik_{zm} \hat{z}_i}
                                    } \nonumber \\
                                  &=  i\sqrt{\frac{1}{\varepsilon_0 V}}\int d\bm{r}  \sum_{\mu}
                                    \frac{\omega_{\mu}^{(\rm{pl})}}{\omega_{\rm{pl}}} \hat{\mathcal{Q}}_{\mu}
                                    \bm{\varepsilon}_{\bm{k}_{\parallel},m}e^{i\bm{k}_{\parallel}\cdot\bm{\rho}} e^{ik_{zm} z}  \cdot \hat{\bm{J}}_{\rm{e}}(\bm{r}) \nonumber \\
                                  &=  \frac{1}{\varepsilon_0 \omega_{\rm{pl}}^2 }\int d\bm{r} 
                                    \hat{\bm{J}}_{\rm{p}}(\bm{r})    \cdot \hat{\bm{J}}_{\rm{e}}(\bm{r})   
\end{align}
Thus, it is expressed as the inner product between the plasmon-induced and the individual-excitation-induced current densities.  
Accordingly, the second quantized form of $\hat{H}_{\mathrm{mf},\rm{e-pl}}$ in Eq.~(\ref{mem_hamiltonian_e-pl}) under the rotating wave approximation is given by
\begin{align}
  &\hat{H}_{\mathrm{mf},\rm{e-pl}} \nonumber \\
 =&\sum_{\mu,\nu}
 \blr{\frac{1}{\varepsilon_0 \omega_{\rm{pl}}^2 }\int d\bm{r} 
 \lr{\bm{\mathcal{J}}_{p;\mu} (\bm{r}) \hat{\mathcal{A}}_{\mu} + \rm{H.c.}}
 \lr{\bm{\mathcal{J}}_{e;\nu}(\bm{r})\hat{\mathcal{B}}_{\nu} + \rm{H.c.}}
 }
 \nonumber \\
 \cong &\sum_{\mu,\nu}
 \blr{\frac{1}{\varepsilon_0 \omega_{\rm{pl}}^2 }\int d\bm{r} \bm{\mathcal{J}}^{\ast}_{e;\nu} (\bm{r})\cdot \bm{\mathcal{J}}_{p;\mu}(\bm{r})} \hat{\mathcal{A}}_{\mu}\hat{\mathcal{B}}^{\dagger}_{\nu} 
 + \blr{\frac{1}{\varepsilon_0 \omega_{\rm{pl}}^2 }\int d\bm{r} \bm{\mathcal{J}}_{e;\nu}(\bm{r})\cdot\bm{\mathcal{J}}^{\ast}_{p;\mu} (\bm{r})} \hat{\mathcal{A}}^{\dagger}_{\mu}\hat{\mathcal{B}}_{\nu}\nonumber \\
 = & \sum_{\mu,\nu}
 \blr{
 g^{(L)}_{\mu\nu} \hat{\mathcal{A}}_{\mu}\hat{\mathcal{B}}^{\dagger}_{\nu} 
 + g^{(L) \ast}_{\mu\nu} \hat{\mathcal{A}}^{\dagger}_{\mu}\hat{\mathcal{B}}_{\nu} 
 }
\end{align}
The coefficient $g^{(L)}_{\mu\nu}$ represents the coupling constant for the residual longitudinal interaction between plasmons and electron-hole pairs, and is defined by
\begin{align}
  g^{(L)  }_{\mu\nu} &= \frac{1}{\varepsilon_0 \omega_{\rm{pl}}^2 }\int d\bm{r} \bm{\mathcal{J}}^{\ast}_{e;\nu} (\bm{r})\cdot \bm{\mathcal{J}}_{p;\mu}(\bm{r}) 
  \label{g_L} 
\end{align}
To summarize, the Hamiltonian for a metallic thin film can be expressed in second quantized form as follows:
\begin{align}
  \hat{H'}_{\mathrm{mf}}       =& \hat{H}_{\mathrm{mf},\rm{pl}} + \hat{H}_{\mathrm{mf},\rm{e}} +\hat{H}_{\mathrm{mf},\rm{e-pl}} + \hat{H}_{\mathrm{mf},C'} 
  \label{mem_hamiltonian_final-2ndq} \\
  \hat{H}_{\mathrm{mf},\rm{pl}}  =& \sum_{\mu} \hbar \omega_{\mu}^{(\rm{pl})}\lr{\hat{\mathcal{A}}^{\dagger}_{\mu}\hat{\mathcal{A}}_{\mu}+\frac{1}{2}}
  \label{mem_hamiltonian_pl-2ndq} \\
  \hat{H}_{\mathrm{mf},\rm{e}}   =& \sum_{\nu} \hbar\omega_{\nu}^{(\rm{e-h})}\hat{\mathcal{B}}^{\dagger}_{\nu}\hat{\mathcal{B}}_{\nu} + E_{\mathrm{F}} -\sum_{\mu} \frac{N}{2} {M}^2_{\mu} 
  \label{mem_hamiltonian_e-2ndq} \\
  \hat{H}_{\mathrm{mf},\rm{e-pl}}=&\sum_{\mu,\nu}
 \blr{
 g^{(L)}_{\mu\nu} \hat{\mathcal{A}}_{\mu}\hat{\mathcal{B}}^{\dagger}_{\nu} 
 + g^{(L) \ast}_{\mu\nu} \hat{\mathcal{A}}^{\dagger}_{\mu}\hat{\mathcal{B}}_{\nu} 
 }
  \label{mem_hamiltonian_e-pl-2ndq} \\
  \hat{H}_{\mathrm{mf},C'}       =&   \sum_{|\bm{k}_{\parallel}|>k_c,m}\frac{1}{2}M_{\bm{k}_{\parallel},m}^{2} \lr{\hat{\beta}^{\dagger}_{\bm{k}_{\parallel},m}\hat{\beta}_{\bm{k}_{\parallel},m}-N} 
  \label{mem_hamiltonian_C'-2ndq}
\end{align}
with
\begin{align}
  \hbar \omega_{\mu}^{(\rm{pl})} &= \frac{\hbar \omega_{\rm{pl}}}{\sqrt{2}}\sqrt{1-\lr{-1}^m e^{-|\bm{k}_{\parallel}|d}} \nonumber \\
  \hbar\omega_{\nu}^{(\rm{e-h})} &= \frac{\hbar^2}{2m^{\ast}}\lr{{\bm{k}}_{\parallel}^2+2\bm{K}_{\parallel}\cdot{\bm{k}}_{\parallel}}+\frac{\hbar^2}{2m^{\ast}}\lr{\frac{\pi}{d}}^2\lr{{n'}^2-n^2} \nonumber \\
   g^{(L)  }_{\mu\nu} &= \frac{1}{\varepsilon_0 \omega_{\rm{pl}}^2 }\int d\bm{r} \bm{\mathcal{J}}^{\ast}_{e;\nu} (\bm{r})\cdot \bm{\mathcal{J}}_{p;\mu}(\bm{r})  \nonumber \\
       \bm{\mathcal{J}}_{p;\mu} (\bm{r}) &= 
    i \varepsilon_0  \omega_{\rm{pl}} \sqrt{ \frac{\hbar \omega_{\mu}^{(\rm{pl})}}{2\varepsilon_0 V}}
  \bm{\varepsilon}_{\bm{k}_{\parallel},m}e^{i\bm{k}_{\parallel}\cdot\bm{\rho}} e^{ik_{zm} z} \nonumber \\
     \bm{\mathcal{J}}_{e;\nu}(\bm{r}) &= 
  \frac{\hbar e}{m^{\ast} V} 
   \blr{
   \begin{matrix}
       \lr{2\bm{K}_{\parallel} + \bm{k}_{\parallel}} \sin k_{z,n}z \sin k_{z,n'}z \\
        \frac{k_{zn'}-k_{zn}}{2i}\sin\lr{k_{zn'}+k_{zn}}z -\frac{k_{zn'}+k_{zn}}{2i}\sin\lr{k_{zn'}-k_{zn}}z  
   \end{matrix}
    } e^{i\bm{k}_{\parallel}\cdot\bm{\rho}} \nonumber 
\end{align}

The term $\hat{H}_{\mathrm{mf},\rm{e-pl}}$ in Eq.~(\ref{mem_hamiltonian_e-pl-2ndq}) describes the residual longitudinal interaction between plasmons and individual excitations.  
In the bulk system, it has been discussed by Pines et al.~\cite{bohm1951collective1,pines1952collective2,bohm1953collective3,pines1953collective4}  
that a unitary transformation eliminating this interaction yields the effective mass of the electron and the spatial dispersion of the plasmon.  
Furthermore, Yamada~\cite{yamada1960mechanism} showed that Landau damping arises from real processes that conserve energy and cannot be removed by such transformations.

The term $\hat{H}_{\mathrm{mem},C'}$ in Eq.~(\ref{mem_hamiltonian_C'-2ndq}) represents the short-range component of the electron–electron Coulomb interaction (i.e., the screening effect).  
While this term generally describes a two-body interaction and is therefore difficult to solve exactly, its short-range nature allows one to treat it phenomenologically as a scattering process, and thus it can be absorbed into a relaxation rate $\Gamma$ or similar parameter. Hence, we neglect it in the present treatment.

\section{Interaction Hamiltonian between matter and radiation field }
Next, we consider introducing Interaction Hamiltonian between matter and radiation field. Since the momentum of an electron in radiation field is transformed as $\hat{\bm{p}}_i \rightarrow \hat{\bm{p}}_i-e\bm{A}(\hat{\bm{r}_i})$, in addition, momentum operator is unitarily transformed by Eq.~(\ref{Unitary_trans_ele_momentum}) therefore, the total current density is written in terms of the following three components:
\begin{align}
  \hat{\bm{J}}(\bm{r}) &= \hat{\bm{J}}_{\rm{p}}(\bm{r}) + \hat{\bm{J}}_{\rm{{e}}}(\bm{r}) + \hat{\bm{J}}_{\rm{{D}}}(\bm{r})
  \label{totalJ_rad} 
\end{align}
$\hat{\bm{J}}_{\rm{p}}$ and $\hat{\bm{J}}_{\rm{{e}}}$ are plasmonic and individual components defined by Eq.~(\ref{Jp}) and Eq.~(\ref{Je}), then $\hat{\bm{J}}_{\rm{{D}}}$ is the Drude components:
\begin{align}
  \hat{\bm{J}}_{\rm{{D}}}(\bm{r}) &= \sum_{i=1}^{N}\blr{-\frac{e^2}{m^{\ast}}\delta(\bm{r}-\hat{\bm{r}}_i)\bm{A}(\hat{\bm{r}}_i)}
  \label{JD} 
\end{align}
Under the Coulomb gauge ($\bm{\nabla}\cdot\bm{A} = 0$), interaction Hamiltonian between matter and radiation field is described in the form of the inner product of the vector potential ($\bm{A}$) and the current density ($\bm{J}$), or the inner product of the transverse electric field ($\bm{E}_{T}$) and the polarization ($\bm{P}$), as follows:
\begin{align}
    \hat{H}_{\mathrm{int}}(t) = - \int d\bm{r} \hat{\bm{J}}(\bm{r},t)\cdot\bm{A}(\bm{r},t) =  - \int d\bm{r} \hat{\bm{P}}(\bm{r},t)\cdot\bm{E}_{T}(\bm{r},t)
\end{align}
note that the longitudinal field interaction, described by the product of the scalar potential ($\phi$) and charge density ($\rho$) and the inner product of the longitudinal electric field ($\bm{E}_{L}$) and polarization ($\bm{P}$) , has already been solved as the exchange interaction between electrons and holes in the matter hamiltonian and is therefore excluded to prevent double counting. 
\subsection{ Constitutive equation }
In the following, the form of the inner product of the transverse electric field and the polarization is adopted as the interaction Hamiltonian.
We assume a monochromatic field, $\bm{E}_{T}(\bm{r},t) = \bm{E}_{T}(\bm{r},\omega)e^{-i\omega t}$ and only consider $\omega$ Fourier components of current density and polarization, $\hat{\bm{J}}(\bm{r},t) = \hat{\bm{J}}(\bm{r},\omega) e^{-i\omega t}, \hat{\bm{P}}(\bm{r},t) = \hat{\bm{P}}(\bm{r},\omega) e^{-i\omega t}$.
From the relation between current density and polarization ($ \hat{\bm{J}}(\bm{r},t) = \frac{\partial}{\partial t} \hat{\bm{P}}(\bm{r},t) $) and Eq.~(\ref{Jp}) and Eq.~(\ref{Je}) and Eq.~(\ref{JD}),each component of the polarization operator is written as follows:
\begin{align}
  \hat{\bm{P}}(\bm{r},\omega) &= \hat{\bm{P}}_{\rm{{p}}}(\bm{r},\omega) + \hat{\bm{P}}_{\rm{{e}}}(\bm{r},\omega) + \hat{\bm{P}}_{\rm{{D}}}(\bm{r},\omega)
  \label{total_P} \\
  \hat{\bm{P}}_{\rm{{p}}}(\bm{r},\omega) &= - \frac{Ne^2}{m^{\ast} \omega V } \sqrt{\frac{1}{\varepsilon_0 V}}
  \sum_{\mu}
  \frac{\omega_{\mu}^{(\rm{pl})}}{\omega_{\rm{pl}}} \hat{\mathcal{Q}}_{\mu}
  \bm{\varepsilon}_{\bm{k}_{\parallel},m}e^{i\bm{k}_{\parallel}\cdot\bm{\rho}} e^{ik_{zm} z}
  \label{Pp} \\
  \hat{\bm{P}}_{\rm{e}}(\bm{r},\omega)  &= \frac{i}{\omega}\sum_{i=1}^{N} \blr{\frac{e\clr{\hat{\bm{p}}_i\delta(\bm{r}-\hat{\bm{r}}_i)+\delta(\bm{r}-\hat{\bm{r}}_i)\hat{\bm{p}}_i}}{2m^{\ast}}}
  \label{Pe} \\
  \hat{\bm{P}}_{\rm{{D}}}(\bm{r},\omega) &= \frac{i}{\omega}\sum_{i=1}^{N}\blr{-\frac{e^2}{m^{\ast}}\delta(\bm{r}-\hat{\bm{x}}_i) \frac{1}{i \omega}\bm{E}_{T}(\hat{\bm{x}}_i,\omega)} = \varepsilon_0 \chi_D \bm{E}_{T}(\bm{r},\omega)
  \label{PD}   
  \end{align}
The 3rd term $\hat{\bm{P}}_{\rm{{D}}}$ corresponds to a Drude response and provides a local Drude susceptibility to the transverse field defined as
\begin{align}
\chi_D  = -\frac{\omega_{\rm{pl}}^2}{\omega^2}
\end{align}
The 1st and 2nd term $\hat{\bm{P}}_{\rm{{p}}},\hat{\bm{P}}_{\rm{{e}}}$ correspond to plasmonic and individual components of polarization. 
The resonant polarization, defined as $\hat{\bm{P}}_{\mathrm{res}} = \hat{\bm{P}}_{\mathrm{p}} + \hat{\bm{P}}_{\mathrm{e}}$, cannot be described by a local susceptibility as in the case of $\hat{\bm{P}}_{\mathrm{D}}$, and thus requires a nonlocal treatment. Within the regime of linear response and under the consideration of the Kubo formula \cite{kubo1957statistical}, the constitutive relation is expressed using the nonlocal susceptibility $\bar{\chi}$ as follows \cite{cho2003optical}:

\begin{align}
    \alr{\hat{\bm{P}}_{\rm{{res}}}(\bm{r},\omega)} &= \int d\bm{r}' \bar{\chi}(\bm{r},\bm{r}',\omega) \bm{E}_{T}(\bm{r}',\omega) 
    \label{constitutive_eq} \\
    \bar{\chi}(\bm{r},\bm{r}',\omega) &= \sum_{\mu}\blr{\frac{\bm{\mathcal{P}}_{p;\mu}(\bm{r},\omega)\bm{\mathcal{P}}^{\ast}_{p;\mu}(\bm{r}',\omega)}{\hbar\omega_{\mu}-\hbar\omega-i\hbar\gamma}} + \sum_{\nu}\blr{\frac{\bm{\mathcal{P}}_{e;\nu}(\bm{r},\omega)\bm{\mathcal{P}}^{\ast}_{e;\nu}(\bm{r}',\omega)}{\hbar\omega_{\nu}-\hbar\omega-i\hbar\gamma}}
\end{align}
Here, for simplicity, the anti-resonance term is neglected and $\bm{\mathcal{P}}_{p;\mu}(\bm{r},\omega)$ and $\bm{\mathcal{P}}_{e;\nu}(\bm{r},\omega)$ are the matrix elements of each polarization oparator:
\begin{align}
    \bm{\mathcal{P}}_{p;\mu}(\bm{r},\omega) &= - \varepsilon_0  \frac{\omega_{\rm{pl}} }{\omega}  \sqrt{ \frac{\hbar \omega_{\mu}^{(\rm{pl})}}{2\varepsilon_0 V}}
  \bm{\varepsilon}_{\bm{k}_{\parallel},m}e^{i\bm{k}_{\parallel}\cdot\bm{\rho}} e^{ik_{zm} z} 
  \label{P_p_mu} \\
   \bm{\mathcal{P}}_{e;\nu}(\bm{r},\omega) &=   \frac{\hbar e}{m^{\ast} \omega V} 
   \blr{
   \begin{matrix}
       i \lr{2\bm{K}_{\parallel} + \bm{k}_{\parallel}} \sin k_{z,n}z \sin k_{z,n'}z \\
        \frac{k_{zn'}-k_{zn}}{2}\sin\lr{k_{zn'}+k_{zn}}z -\frac{k_{zn'}+k_{zn}}{2}\sin\lr{k_{zn'}-k_{zn}}z  
   \end{matrix}
    } e^{i\bm{k}_{\parallel}\cdot\bm{\rho}}
\end{align}
\subsection{Microscopic Maxwell equation }
If the polarization $\bm{P}(\bm{r})$ exists in vacuum, the transverse electric field $\bm{E}_{\rm{T}}(\bm{r},\omega)$ satisfies the following equation by Maxwell's equations in vacuum:
\begin{align}
    \blr{\bm{\nabla}^2 + \frac{\omega^2}{c^2} }\bm{E}_{\rm{T}} (\bm{r},\omega) = -\mu_0 \omega^2 \bm{P}(\bm{r},\omega)
\end{align}
From Eq.~(\ref{total_P}), the total polarization can be written as $\bm{P} = {\bm{P}}_{\mathrm{res}} + {\bm{P}}_{\mathrm{D}}$. The local component ${\bm{P}}_{\mathrm{D}}$, which can be expressed in the form of a local response, can be renormalized into the background dielectric constant $\varepsilon_{\rm{bg}}(\bm{r})$. After this renormalization, Microscopic Maxwell equation becomes:
\begin{align}
    \blr{\bm{\nabla}^2 + \frac{\omega^2}{c^2} \varepsilon_{\rm{bg}}(\bm{r}) }\bm{E}_{\rm{T}} (\bm{r},\omega) &= -\mu_0 \omega^2 \bm{P}_{\rm{res}}(\bm{r},\omega) 
    \label{M-eq_2}\\
    \varepsilon_{\rm{bg}}(\bm{r}) &= 1+\chi_{D}\Theta(\bm{r}) 
\end{align}
where $\Theta(\bm{r})=1$or $\Theta(\bm{r})=0$, depending on whether the point $\bm{r}$ is inside or outside the thin film.
The solution of Eq.~(\ref{M-eq_2}) can be expressed as follows:
\begin{align}
  \bm{E}_T(\bm{r},\omega) &= \bm{E}_{T,\rm{bg}} (\bm{r},\omega) + \frac{1}{\varepsilon_0} \int d\bm{r}' \bar{\bm{G}}_{T,\rm{mem}} (\bm{r},\bm{r}',\omega) {\bm{P}}_{\rm{{res}}}(\bm{r}',\omega)
  \label{maxwell-eq_bg}
\end{align}
using the background electric field $\lr{\bm{E}_{T,\rm{bg}}}$ and dyadic Green's function for Radiation field in a film structure $\lr{\bar{\bm{G}}_{T,\rm{mem}}}$ satisfying the following equation:
\begin{align}
    \blr{\bm{\nabla}^2 + \frac{\omega^2}{c^2} \varepsilon_{\rm{bg}}(\bm{r}) }\bm{E}_{T,\rm{bg}}(\bm{r},\omega) &= 0 
    \label{E_bg} \\
    \blr{\bm{\nabla}^2 + \frac{\omega^2}{c^2} \varepsilon_{\rm{bg}}(\bm{r}) } \bar{\bm{G}}_{T,\rm{mem}} (\bm{r},\bm{r}',\omega) &= - \frac{\omega^2}{c^2}\delta(\bm{r}-\bm{r}')\bar{\mathbb{I}} 
\end{align}
The explicit form of the dyadic Green's function for Radiation field in a film structure $\lr{\bar{\bm{G}}_{T,\rm{mem}}}$ is described in detail in Ref.~\cite{chew1999waves}, and the present study also makes use of that expression.

\subsection{Self-consistent equation }
Substituting Eq.~(\ref{constitutive_eq}) into Eq.~(\ref{maxwell-eq_bg}),
\begin{align}
   \bm{E}_T(\bm{r},\omega) &= \bm{E}_{T,\rm{bg}} (\bm{r},\omega) + \frac{1}{\varepsilon_0} \iint d\bm{r}' d\bm{r}'' \bar{\bm{G}}_{T,\rm{mem}} (\bm{r},\bm{r}',\omega) \bar{\chi}(\bm{r}',\bm{r}'',\omega) \bm{E}_T(\bm{r}'',\omega)
\end{align}
Multiply both sides by $\bm{\mathcal{P}}^{\ast}_{p;\mu}(\bm{r},\omega)$ and integrate by $\bm{r}$,
\begin{align}
   &\int d\bm{r} \bm{\mathcal{P}}^{\ast}_{p;\mu}(\bm{r},\omega)\cdot\bm{E}_T(\bm{r},\omega) = \int d\bm{r} \bm{\mathcal{P}}^{\ast}_{p;\mu}(\bm{r},\omega)\cdot \bm{E}_{T,\rm{bg}} (\bm{r},\omega) \nonumber \\
   &+ \frac{1}{\varepsilon_0} \iiint d\bm{r} d\bm{r}' d\bm{r}'' \bm{\mathcal{P}}^{\ast}_{p;\mu}(\bm{r},\omega) \bar{\bm{G}}_{T,\rm{mem}} (\bm{r},\bm{r}',\omega) \blr{\sum_{\mu'}\clr{\frac{\bm{\mathcal{P}}_{p;\mu'}(\bm{r}',\omega)\bm{\mathcal{P}}^{\ast}_{p;\mu'}(\bm{r}'',\omega)}{\hbar\omega_{\mu'}-\hbar\omega-i\hbar\gamma}} + \sum_{\nu'}\clr{\frac{\bm{\mathcal{P}}_{e;\nu'}(\bm{r}',\omega)\bm{\mathcal{P}}^{\ast}_{e;\nu'}(\bm{r}'',\omega)}{\hbar\omega_{\nu'}-\hbar\omega-i\hbar\gamma}}} \bm{E}_T(\bm{r}'',\omega)
   \label{SC_1}
\end{align}
and 
Multiply both sides by $\bm{\mathcal{P}}^{\ast}_{e;\nu}(\bm{r},\omega)$ and integrate by $\bm{r}$,
\begin{align}
   &\int d\bm{r} \bm{\mathcal{P}}^{\ast}_{e;\nu}(\bm{r},\omega)\cdot\bm{E}_T(\bm{r},\omega) = \int d\bm{r} \bm{\mathcal{P}}^{\ast}_{e;\nu}(\bm{r},\omega)\cdot \bm{E}_{T,\rm{bg}} (\bm{r},\omega) \nonumber \\
   &+ \frac{1}{\varepsilon_0} \iiint d\bm{r} d\bm{r}' d\bm{r}'' \bm{\mathcal{P}}^{\ast}_{e;\nu}(\bm{r},\omega) \bar{\bm{G}}_{T,\rm{mem}} (\bm{r},\bm{r}',\omega) \blr{\sum_{\mu'}\clr{\frac{\bm{\mathcal{P}}_{p;\mu'}(\bm{r}',\omega)\bm{\mathcal{P}}^{\ast}_{p;\mu'}(\bm{r}'',\omega)}{\hbar\omega_{\mu'}-\hbar\omega-i\hbar\gamma}} + \sum_{\nu'}\clr{\frac{\bm{\mathcal{P}}_{e;\nu'}(\bm{r}',\omega)\bm{\mathcal{P}}^{\ast}_{e;\nu'}(\bm{r}'',\omega)}{\hbar\omega_{\nu'}-\hbar\omega-i\hbar\gamma}}} \bm{E}_T(\bm{r}'',\omega)
   \label{SC_2}
\end{align}
Defining $h_{\mu(\nu)}, F_{\mu}, F_{\nu}, F^{(\mathrm{b})}{\mu}, F^{(\mathrm{b})}{\nu}, Z_{\mu\mu'}, g^{(T)}{\mu\nu},$ and $Y{\nu\nu'}$ as follows:
\begin{align}
    h_{\mu(\nu)} &= \frac{1}{\hbar\omega_{\mu(\nu)}-\hbar\omega -i\hbar\gamma} 
    \label{h_munu}\\
    F_{\mu}      &= \int d\bm{r} \bm{\mathcal{P}}^{\ast}_{p;\mu}(\bm{r},\omega)\cdot\bm{E}_T(\bm{r},\omega) 
    \label{F_mu}\\
    F_{\nu}      &= \int d\bm{r} \bm{\mathcal{P}}^{\ast}_{e;\nu}(\bm{r},\omega)\cdot\bm{E}_T(\bm{r},\omega) 
    \label{F_nu}\\
    F^{(\rm{b})}_{\mu} &= \int d\bm{r} \bm{\mathcal{P}}^{\ast}_{p;\mu}(\bm{r},\omega)\cdot \bm{E}_{T,\rm{bg}} (\bm{r},\omega) \label{F_b_mu} \\
    F^{(\rm{b})}_{\nu} &= \int d\bm{r} \bm{\mathcal{P}}^{\ast}_{e;\nu}(\bm{r},\omega)\cdot \bm{E}_{T,\rm{bg}} (\bm{r},\omega) \label{F_b_nu} \\
    Z_{\mu\mu'} &= -\frac{1 }{\varepsilon_0} \iint d\bm{r}d\bm{r}'  \bm{\mathcal{P}}^{\ast}_{p;\mu}(\bm{r},\omega) \bar{\bm{G}}_{T,\rm{mem}} (\bm{r},\bm{r}',\omega) \bm{\mathcal{P}}_{p;\mu'}(\bm{r}',\omega) 
    \label{Z_mumu'} \\
    g^{(T)}_{\mu\nu} &= -\frac{1 }{\varepsilon_0} \iint d\bm{r}d\bm{r}'  \bm{\mathcal{P}}^{\ast}_{p;\mu}(\bm{r},\omega) \bar{\bm{G}}_{T,\rm{mem}} (\bm{r},\bm{r}',\omega) \bm{\mathcal{P}}_{e;\nu}(\bm{r}',\omega) 
    \label{gT_munu} \\
    Y_{\nu\nu'} &= -\frac{1 }{\varepsilon_0} \iint d\bm{r}d\bm{r}'  \bm{\mathcal{P}}^{\ast}_{e;\nu}(\bm{r},\omega) \bar{\bm{G}}_{T,\rm{mem}} (\bm{r},\bm{r}',\omega) \bm{\mathcal{P}}_{e;\nu'}(\bm{r}',\omega) 
    \label{Y_nunu'} 
\end{align}
Eq.~(\ref{SC_1}) and Eq.~(\ref{SC_2}) can be written as follows:
\begin{align}
    \sum_{\mu'}\blr{\lr{\hbar\omega_{\mu'}-\hbar\omega-i\hbar\gamma}\delta_{\mu\mu'} + Z_{\mu\mu'}}\blr{h_{\mu'}F_{\mu'}} + \sum_{\nu'} \blr{g^{(T)}_{\mu\nu'}}\blr{h_{\nu'}F_{\nu'}} &= F_{\mu}^{(\rm{b})} 
    \label{SC_1_Re} \\
    \sum_{\mu'}\blr{g^{(T)}_{\mu'\nu}}\blr{h_{\mu'}F_{\mu'}} + \sum_{\nu'}\blr{\lr{\hbar\omega_{\nu'}-\hbar\omega-i\hbar\gamma}\delta_{\nu\nu'} + Y_{\nu\nu'}}\blr{h_{\nu'}F_{\nu'}} &= F_{\nu}^{(\rm{b})} 
    \label{SC_2_Re}     
\end{align}
Accordingly, by expressing these equations in matrix form, we obtain the following set of linear equations (self-consistent equations), which determine the response field $\bm{E}_{T}$ and the corresponding resonant polarization $\bm{P}_{\rm{res}}$.
\begin{align}
    \begin{bmatrix}
      \bar{\bm{S}}        & \bar{\bm{g}}^{\rm{Tr}} \\
      \bar{\bm{g}}        & \bar{\bm{T}}
    \end{bmatrix}
    \begin{bmatrix}
     \bm{F}_\mathrm{p} \\
     \bm{F}_\mathrm{e}
    \end{bmatrix} =     
    \begin{bmatrix}
     \bm{F}^{(\mathrm{b})}_\mathrm{p} \\
     \bm{F}^{(\mathrm{b})}_\mathrm{e}
    \end{bmatrix}
    \label{SC_Eq}
\end{align}

\begin{align}
    \bar{\bm{S}}_{\mu\mu'} &= \lr{\hbar\omega_{\mu'}-\hbar\omega-i\hbar\gamma}\delta_{\mu\mu'} + Z_{\mu\mu'}
    \label{S_mumu'} \\
    \bar{\bm{g}}_{\mu\nu}  &= g^{(T)}_{\mu\nu}
    \label{gT_munu} \\
    \bar{\bm{T}}_{\nu\nu'} &= \lr{\hbar\omega_{\nu'}-\hbar\omega-i\hbar\gamma}\delta_{\nu\nu'} + Y_{\nu\nu'}
    \label{T_nunu'} \\
    \bm{F}_{\rm{p}} &= \blr{\cdots,h_{\mu}F_{\mu},\cdots}^{\rm{Tr}}
    \label{Fp} \\
    \bm{F}_{\rm{e}} &= \blr{\cdots,h_{\nu}F_{\nu},\cdots}^{\rm{Tr}}
    \label{Fe} \\
    \bm{F}_{\rm{p}}^{(\rm{b})} &= \blr{\cdots,F_{\mu}^{(\rm{b})},\cdots}^{\rm{Tr}}
    \label{Fpb} \\
    \bm{F}_{\rm{e}}^{(\rm{b})} &= \blr{\cdots,F_{\nu}^{(\rm{b})},\cdots}^{\rm{Tr}}
    \label{Feb} 
    \end{align}
Even in the absence of an incident field, i.e., when 
$
\begin{bmatrix}
 \bm{F}^{(\mathrm{b})}_\mathrm{p} \\
 \bm{F}^{(\mathrm{b})}_\mathrm{e}
\end{bmatrix}
= \bm{0}
$
in Eq.~(\ref{SC_Eq}), 
a solution exists under the following condition:
\begin{align}
    \rm{Det} \blr{
    \begin{matrix}
      \bar{\bm{S}}        & \bar{\bm{g}}^{\rm{Tr}} \\
      \bar{\bm{g}}        & \bar{\bm{T}}
    \end{matrix}
    } = 0 
\end{align}
which determines the self-sustained modes of the matter–radiation coupled system. 
By analyzing these self-sustained modes, we can calculate the dispersion relation, group velocity, 
and radiative linewidth.

\section{Dispersion relation of Surface Plasmon Polariton }
To confirm that our formalism encompasses the conventional optical response theory based on the Drude model, we demonstrate that the dispersion relation of surface plasmon polaritons can be derived by neglecting the degrees of freedom associated with individual excitations. By neglecting the degrees of freedom associated with individual excitations, the self-sustained modes of the coupled system of plasmons and the radiation field are given by the following equation :
\begin{align}
    \rm{Det} \blr{\bar{\bm{S}}} = 0
\end{align}
Using the free electron density of silver and solving the above equations for various film thickness $d$, the following dispersion relations are obtained (shown in FIG.1).
\begin{figure}[htbp]
  \centering
  \begin{subfigure}{0.3\textwidth}
    \includegraphics[width=\linewidth]{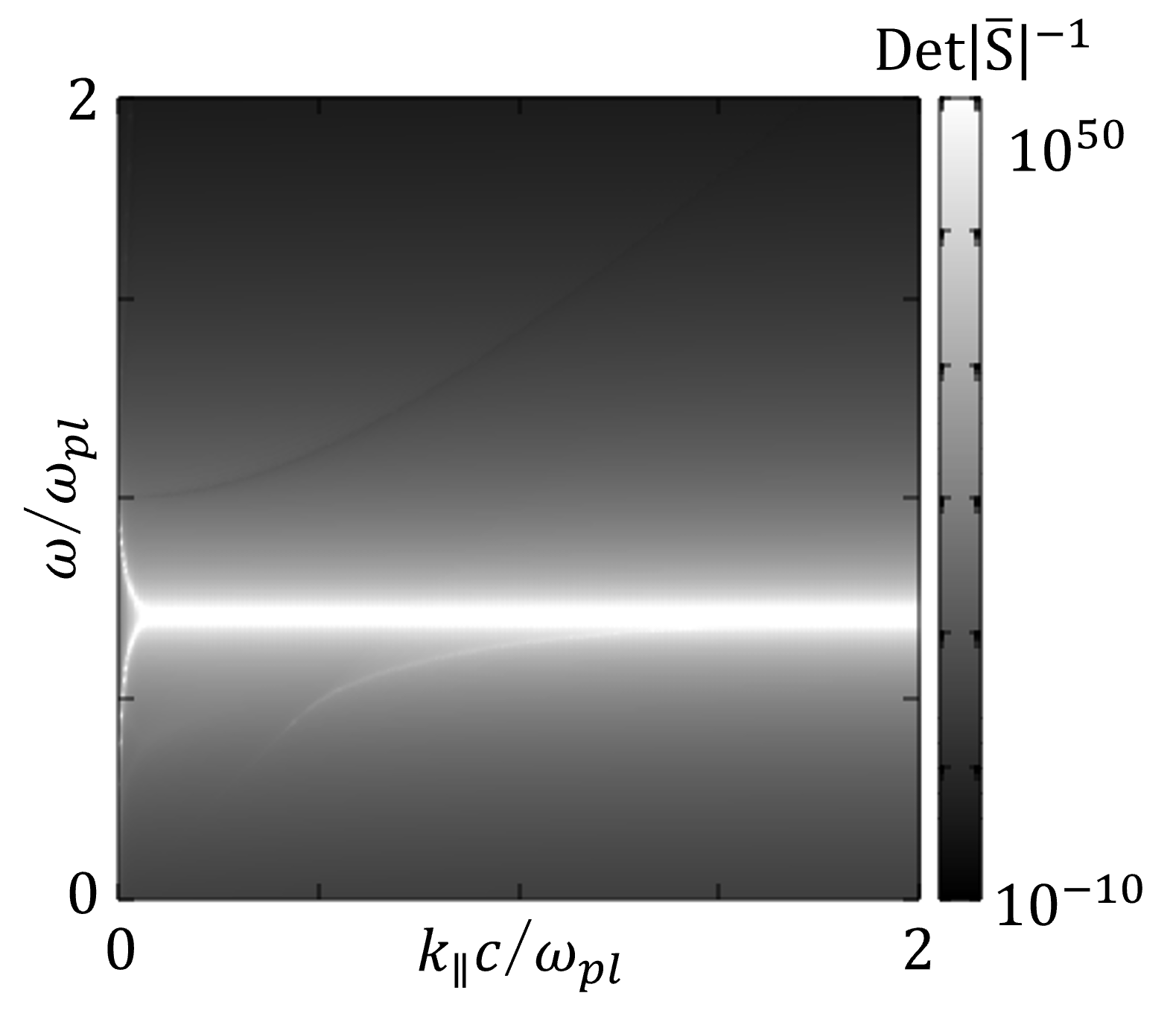}
    \caption{thickness 1000nm}
  \end{subfigure}
  \hfill
  \begin{subfigure}{0.3\textwidth}
    \includegraphics[width=\linewidth]{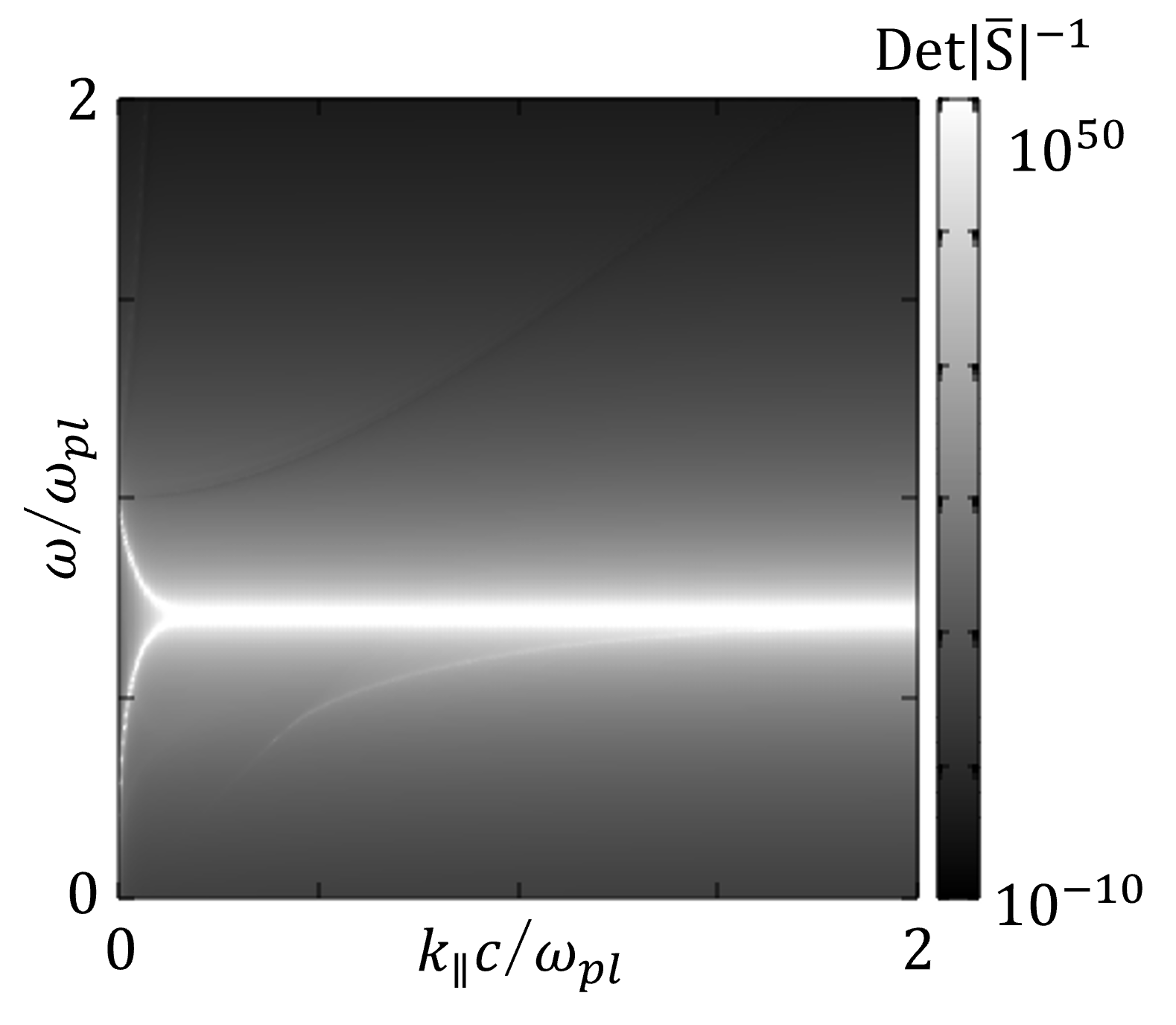}
    \caption{thickness 500nm}
  \end{subfigure}
  \hfill
  \begin{subfigure}{0.3\textwidth}
    \includegraphics[width=\linewidth]{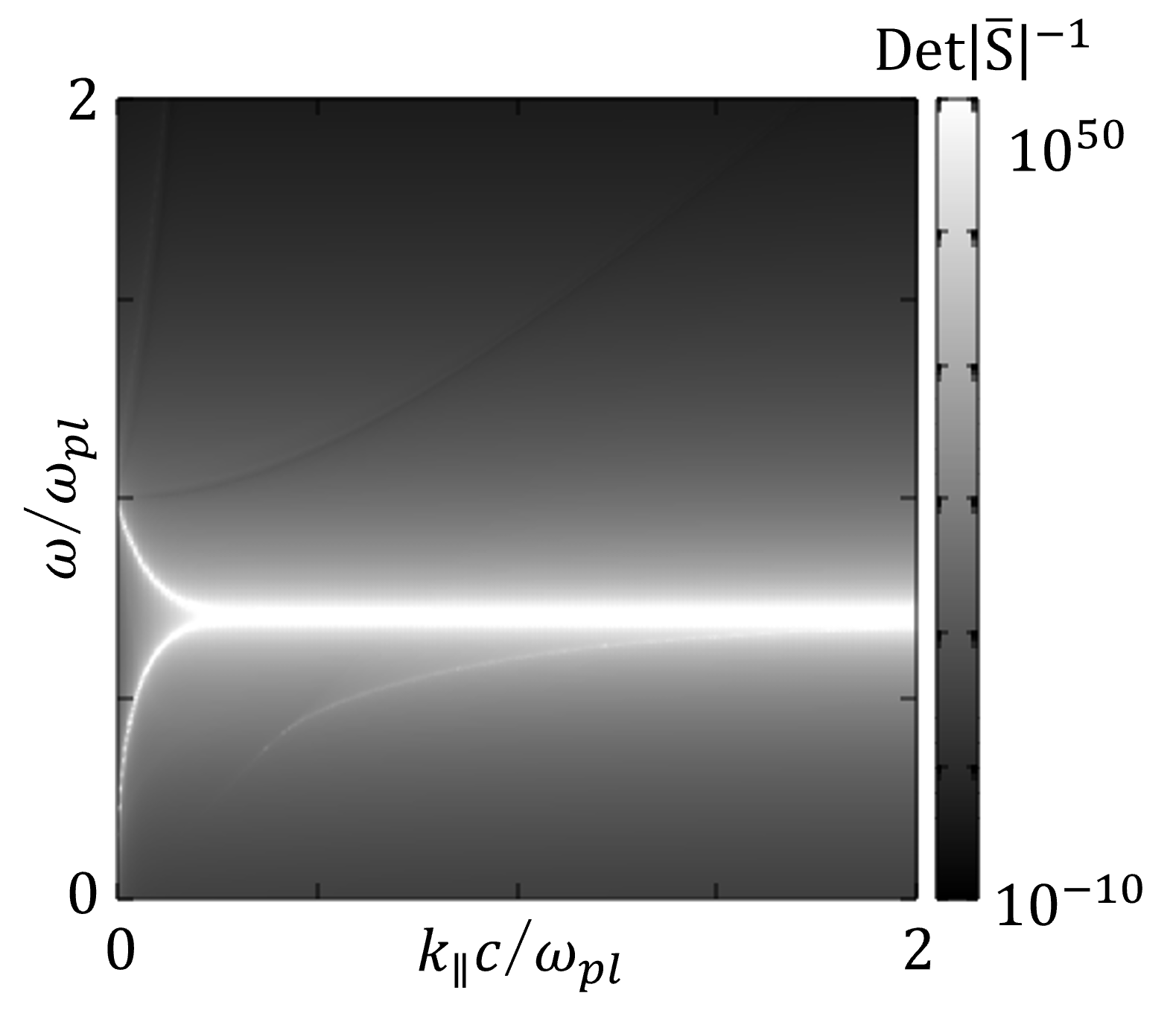}
    \caption{thickness 300nm}
  \end{subfigure}

  \vspace{1em}

  \begin{subfigure}{0.3\textwidth}
    \includegraphics[width=\linewidth]{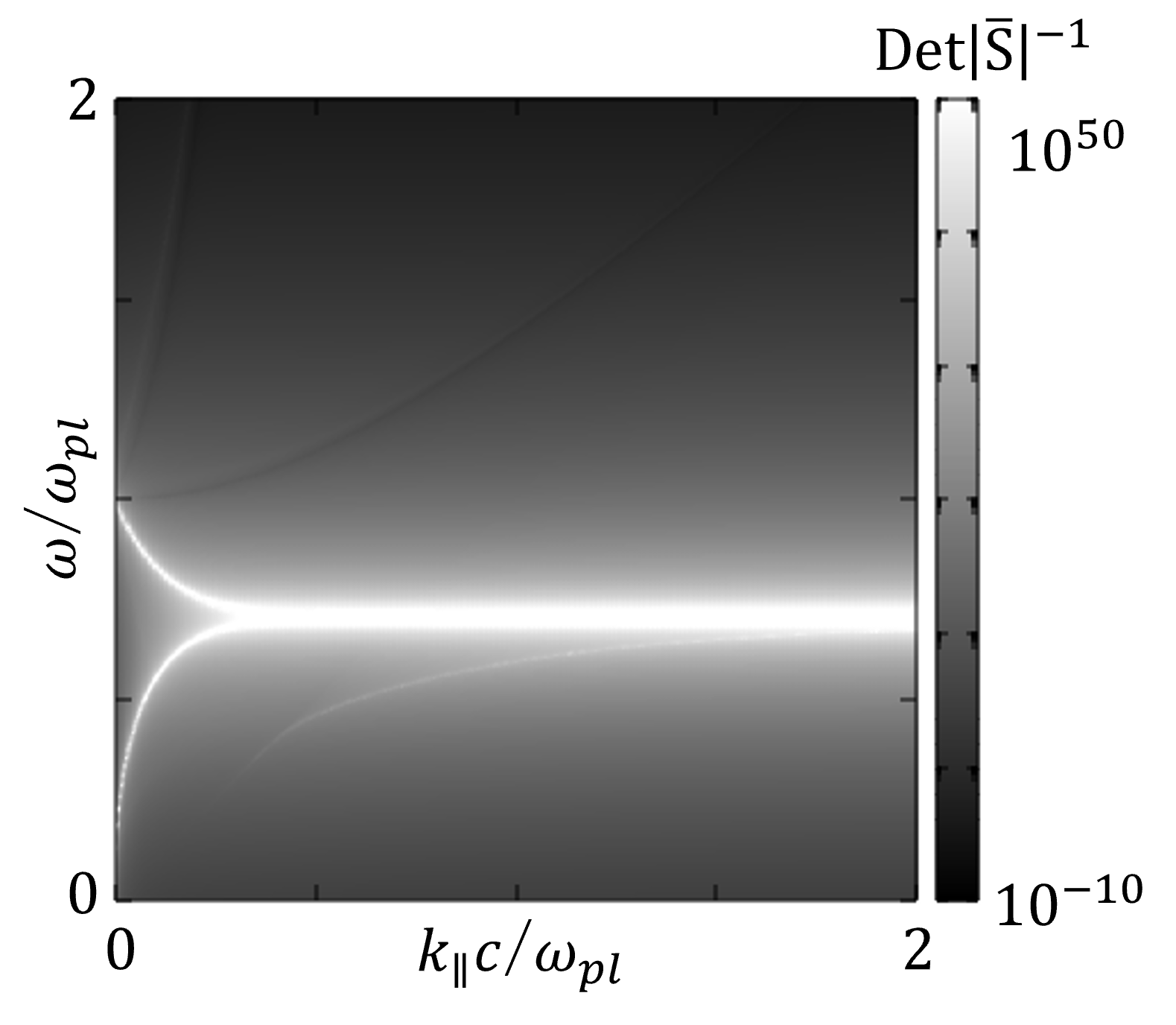}
    \caption{thickness 100nm}
  \end{subfigure}
  \hfill
  \begin{subfigure}{0.3\textwidth}
    \includegraphics[width=\linewidth]{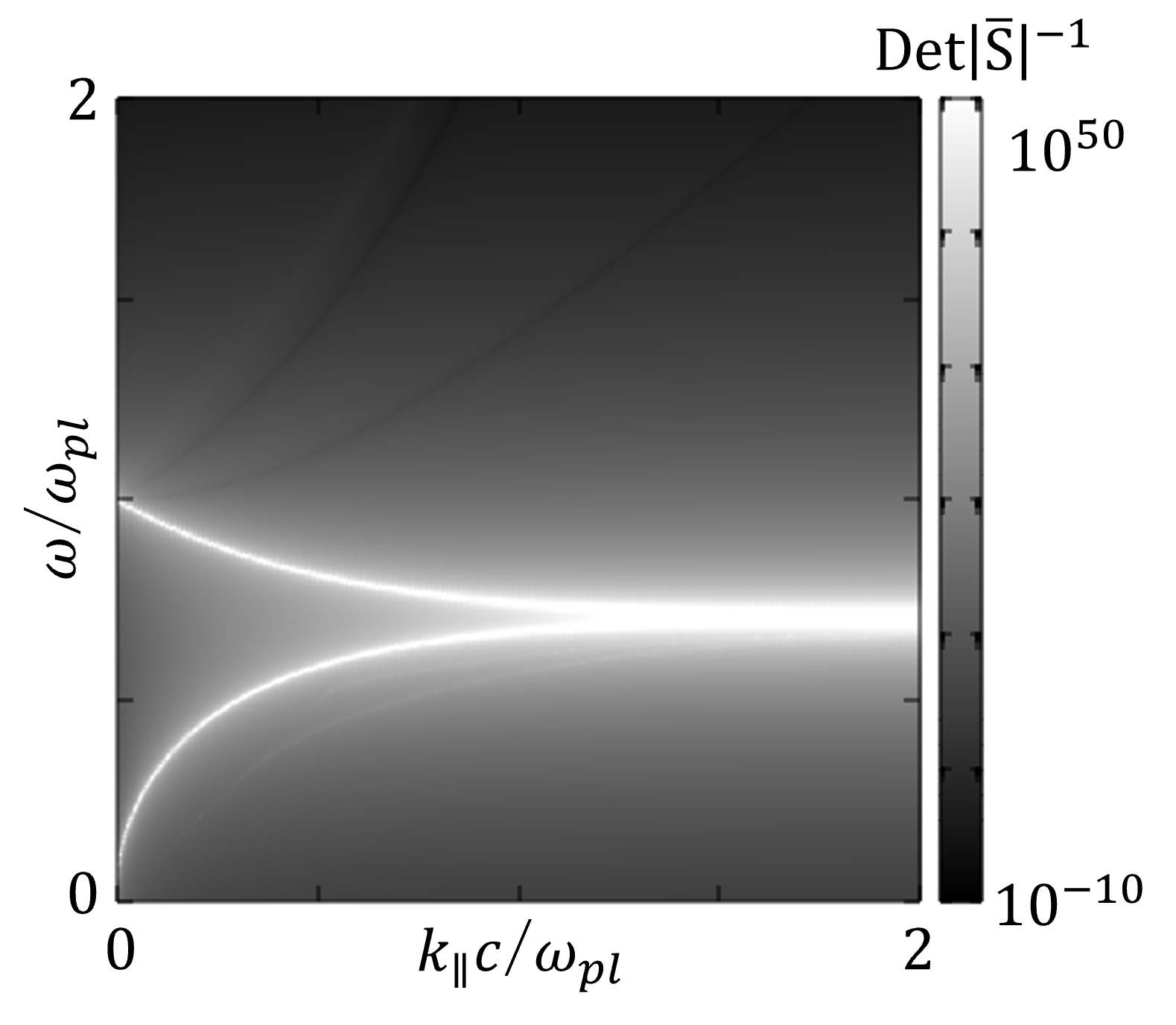}
    \caption{thickness 50nm}
  \end{subfigure}
  \hfill
  \begin{subfigure}{0.3\textwidth}
    \includegraphics[width=\linewidth]{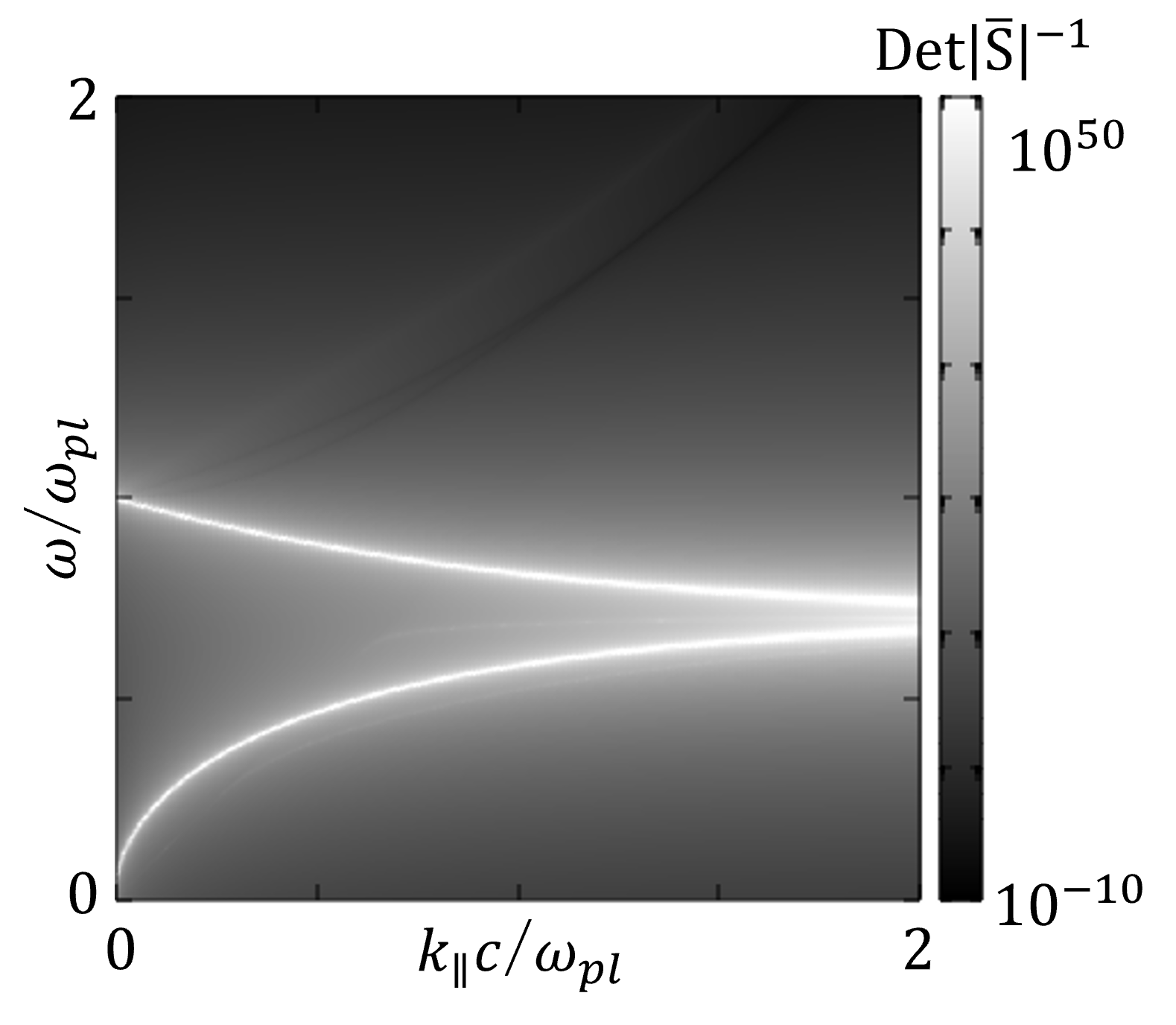}
    \caption{thickness 25nm}
  \end{subfigure}
  \caption{Dispersion relations of surface plasmon polaritons in Ag thin films with varying thicknesses}
\end{figure}
Here, we consider plasmon modes with indices $m \sim -10$ to $10$ and plot the inverse of $\det|\bar{\bm{S}}|$ as a function of each $(k_{\parallel}, \omega)$, where each matrix elements of $\bar{\bm{S}}$ is normalized by $\hbar\omega_{\mathrm{pl}}$. For each film thickness $d$, the thick and bright lines represent modes that do not couple to the radiation field and closely follow the intrinsic plasmon dispersion $\omega_{\mu}^{(\rm{pl})}$. In contrast, we observe dispersive branches that asymptotically approach the light line, which corresponds to the dispersion relation $\omega = ck_{\parallel}$ (a line with unit slope in the $\omega$–$k_{\parallel}$ plane). These branches represent the surface plasmon polariton (SPP) modes. Furthermore, in thicker films, the SPP modes are nearly degenerate, whereas in thinner films, they split into long-range and short-range propagating SPPs due to interference between the top and bottom surfaces. This behavior is consistent with conventional plasmon theory based on the Drude model~\cite{economou1969surface}, demonstrating that our theoretical framework encompasses earlier models that neglect the response of individual excitations.

\newpage

\section{Discussions}

Bohm and Pines~\cite{bohm1951collective1,pines1952collective2,bohm1953collective3,pines1953collective4} formulated the electronic states in bulk metals using a basis composed of collective excitations (plasmons) and individual excitations (electron--hole pairs). Subsequently, Yamada~\cite{yamada1960mechanism} demonstrated that the residual longitudinal interaction between these two excitation types leads to Landau damping within this theoretical framework. Modern theories of hot carrier generation~\cite{sundararaman2014theoretical,manjavacas2014plasmon,besteiro2017understanding,zhang2021theory} are largely based on this framework, treating the generation of carriers from plasmons as a phenomenological relaxation process. However, this description is valid only in bulk systems, where transverse electromagnetic fields and longitudinal plasmon modes remain orthogonal and thus do not interact. In contrast, finite systems with surfaces allow coupling to the radiation field, necessitating an extension of the Bohm--Pines theory to explicitly include the light--matter interaction Hamiltonian.

In this study, we extend the Bohm--Pines framework to an infinitely extended metallic thin film and formulate a generalized theory of the optical response of metals. This theory self-consistently incorporates the light--matter interaction and includes quantum effects. As a result, we demonstrate that the total Hamiltonian can be expressed in a form that includes radiative corrections:
\begin{align}
  \hat{H}_{\mathrm{tot}}       =& \hat{H}_{\mathrm{mf},\rm{pl}} + \hat{H}_{\mathrm{mf},\rm{e}} +\hat{H}_{\mathrm{mf},\rm{e-pl}} + \hat{H}_{\mathrm{mf},C'} 
  \label{mem_hamiltonian_final-2ndq_Rad} \\
  \hat{H}_{\mathrm{mf},\rm{pl}}  =& \sum_{\mu\mu'}\blr{ \hbar \omega_{\mu}^{(\rm{pl})} \delta_{\mu\mu'} + \red{Z_{\mu\mu'}} }\lr{\hat{\mathcal{A}}^{\dagger}_{\mu}\hat{\mathcal{A}}_{\mu'}+\frac{1}{2}}
  \label{mem_hamiltonian_pl-2ndq_Rad} \\
  \hat{H}_{\mathrm{mf},\rm{e}}   =& \sum_{\nu\nu'}\blr{ \hbar\omega_{\nu}^{(\rm{e-h})}\delta_{\nu\nu'} +  \red{Y_{\nu\nu'}}}\hat{\mathcal{B}}^{\dagger}_{\nu}\hat{\mathcal{B}}_{\nu'} + E_{\mathrm{F}} -\sum_{\mu} \frac{N}{2} {M}^2_{\mu} 
  \label{mem_hamiltonian_e-2ndq_Rad} \\
  \hat{H}_{\mathrm{mf},\rm{e-pl}}=&\sum_{\mu,\nu}
 \blr{
 \lr{g^{(L)}_{\mu\nu} + \red{g^{(T)}_{\mu\nu}}}\hat{\mathcal{A}}_{\mu}\hat{\mathcal{B}}^{\dagger}_{\nu} 
 + \lr{g^{(L)\ast}_{\mu\nu} + \red{g^{(T)\ast}_{\mu\nu}}} \hat{\mathcal{A}}^{\dagger}_{\mu}\hat{\mathcal{B}}_{\nu} 
 }
  \label{mem_hamiltonian_e-pl-2ndq_Rad} \\
  \hat{H}_{\mathrm{mf},C'}       =&   \sum_{|\bm{k}_{\parallel}|>k_c,m}\frac{1}{2}M_{\bm{k}_{\parallel},m}^{2} \lr{\hat{\beta}^{\dagger}_{\bm{k}_{\parallel},m}\hat{\beta}_{\bm{k}_{\parallel},m}-N} 
  \label{mem_hamiltonian_C'-2ndq_Rad}
\end{align}
with
\begin{align}
       Z_{\mu\mu'} &= -\frac{1 }{\varepsilon_0} \iint d\bm{r}d\bm{r}'  \bm{\mathcal{P}}^{\ast}_{p;\mu}(\bm{r},\omega) \bar{\bm{G}}_{T,\rm{mem}} (\bm{r},\bm{r}',\omega) \bm{\mathcal{P}}_{p;\mu'}(\bm{r}',\omega) 
    \nonumber \\
    Y_{\nu\nu'} &= -\frac{1 }{\varepsilon_0} \iint d\bm{r}d\bm{r}'  \bm{\mathcal{P}}^{\ast}_{e;\nu}(\bm{r},\omega) \bar{\bm{G}}_{T,\rm{mem}} (\bm{r},\bm{r}',\omega) \bm{\mathcal{P}}_{e;\nu'}(\bm{r}',\omega) 
    \nonumber \\
    g^{(T)}_{\mu\nu} &= -\frac{1 }{\varepsilon_0} \iint d\bm{r}d\bm{r}'  \bm{\mathcal{P}}^{\ast}_{p;\mu}(\bm{r},\omega) \bar{\bm{G}}_{T,\rm{mem}} (\bm{r},\bm{r}',\omega) \bm{\mathcal{P}}_{e;\nu}(\bm{r}',\omega) 
   \nonumber 
\end{align}

By analyzing the radiative correction for the plasmon, represented by $Z_{\mu\mu'}$, we derive the well-known dispersion relation for surface plasmon polaritons~\cite{economou1969surface}, thereby confirming that our theory reproduces existing results in the limit where individual excitations are neglected.

Importantly, our formulation also reveals the existence of radiative coupling between collective and individual excitations, described by the term $g^{(T)}$. This coupling introduces a novel mechanism for hot carrier generation, distinct from the conventional Landau damping process driven by $g^{(L)}$ shown by Yamada~\cite{yamada1960mechanism}. In particular, since $g^{(T)}$ exhibits frequency dependence, it is highly promising as a theoretical foundation for explaining the frequency-dependent behavior of internal quantum efficiency (IQE) observed in hot carrier-based photoelectric conversion~\cite{shi2018enhanced}, as mentioned in the Introduction.

Moreover, if this radiative coupling between collective and individual excitations can be sufficiently enhanced, it may be possible to induce Fano-type resonance~\cite{fano1961effects} spontaneously within the intrinsic excitations of metals —without relying on external quantum systems. Such a mechanism could enable the efficient harvesting of light energy by utilizing plasmonic modes with large absorption cross-sections as antennas, and resonantly generating hot carriers. This approach offers a compelling route toward further advancements in the efficiency of plasmon-induced hot carrier generation.

\newpage

\onecolumngrid
\appendix

\section{Supplemental Details}
\label{appendix_A}
Given the two-dimensional Fourier transform of $f(x,y) = \frac{1}{r} = \frac{1}{\clr{x^2+y^2+z^2}^{\frac{1}{2}}}$ ($z$ is constant.),
\begin{align}
  F(k_x, k_y) 
  &= 2\pi \int_{0}^{\infty} dr \frac{r}{\clr{r^2+z^2}^{\frac{1}{2}}} J_{0} (k_{\parallel} r )
  \nonumber \\
  &= 2\pi H(k_{\parallel},z)
  \end{align}
where $J_{0}$ is the 0th-order of the first kind Bessel function expressed by
\begin{align}
  J_{n}(x) =  \sum_{m=0}^{\infty} \frac{\lr{-1}^m}{m!\lr{m+n}!}\lr{\frac{x}{2}}^{2m+n}
\label{Bessel-function}
\end{align}
and it satisfies the following equation :
\begin{align}
  J_{0}(x) &= \frac{1}{2\pi} \int_{0}^{2\pi} d\theta e^{-ix\cos\theta}
  \label{Bessel's integral} 
\end{align}
$H(k_{\parallel},z)$ is the 0th-order Hankel transform for $h(r,z) = \frac{1}{\clr{r^2+z^2}^{\frac{1}{2}}}$ and is expressed as follows:
\begin{align}
  H(k_{\parallel},z) &=\int_{0}^{\infty} h(r,z) J_{0} (k_{\parallel} r) r dr  
  \nonumber \\
                     &= \frac{e^{-k_{\parallel}|z|} }{k_{\parallel}}
  \label{Hankel-transformation} 
\end{align}
Thus, the partial Fourier integral representation of the three-dimensional Coulomb potential $f(\bm{\rho},z)= \frac{1}{r}$ in the in-plane direction is written as follows:
\begin{align}
  f(\bm{\rho},z) 
  &= \iint \frac{d^2 \bm{k}_{\parallel}}{\lr{2\pi}^2} \frac{2\pi}{|\bm{k}_{\parallel}|} e^{-|\bm{k}_{\parallel}||z|}e^{i\bm{k}_{\parallel}\cdot\bm{\rho}}
  \label{basis_mem_appendix_1_1}
\end{align}
Given a periodic function $ h(z) = e^{-|\bm{k}_{\parallel}||z|} (-d\leq z \leq d) $ that is continuous at $z=\pm d$ and periodic in $2d$, the Fourier coefficients of $h(z)$ are
\begin{align}
  a_0 = \frac{1}{d}\int_{-d}^{d} h(z) dz = \frac{2}{|\bm{k}_{\parallel}|d}\clr{1-e^{-|\bm{k}_{\parallel}|d}}
\end{align}
For $m= 1,2,3,\cdots$, with $k_{zm} = \frac{m \pi}{d}$,
\begin{align}
  a_m = \frac{1}{d}\int_{-d}^{d} h(z)\cos(k_{zm}) dz = \frac{2|\bm{k}_{\parallel}|}{\lr{|\bm{k}_{\parallel}|^2 + k_{zm}^2}d}\clr{1-\lr{-1}^{m}e^{-|\bm{k}_{\parallel}|d}}
\end{align}
\begin{align}
  b_m = \frac{1}{d}\int_{-d}^{d} h(z)\sin(k_{zm}) dz = 0
\end{align}
Therefore, the Fourier series expansion of $h(z)$ is
\begin{align}
  h(z) = \frac{1}{|\bm{k}_{\parallel}|d}\clr{1-e^{-|\bm{k}_{\parallel}|d}} + \frac{2}{d}\sum_{m=1}^{\infty} \frac{|\bm{k}_{\parallel}|}{\lr{|\bm{k}_{\parallel}|^2 + k_{zm}^2}}\clr{1-\lr{-1}^{m}e^{-|\bm{k}_{\parallel}|d}} \cos(k_{zm} z)
\end{align}
The complex Fourier expansion of $h(z)$ is
\begin{align}
  h(z) = \frac{1}{d}\sum_{m=-\infty}^{\infty} \frac{|\bm{k}_{\parallel}|}{\lr{|\bm{k}_{\parallel}|^2 + k_{zm}^2}}\clr{1-\lr{-1}^{m}e^{-|\bm{k}_{\parallel}|d}} e^{ik_{zm} z}
\end{align}
which agrees with Eq. ($\mathrm{A\cdot2}$) in Appendix I of the paper by H. Kanazawa~\cite{kanazawa1961plasma}. \\

Thus, the Fourier integral representation of the three-dimensional Coulomb potential in film system  $f(\bm{\rho},z)= \frac{1}{r}$ is obtained as
\begin{align}
  f(\bm{\rho},z) = \iint\frac{ d^2 \bm{k}_{\parallel} }{\lr{2\pi}^2} \frac{1}{d}\sum_{m=-\infty}^{\infty} \frac{2\pi}{\bm{k}_{\parallel}^2 + k_{zm}^2}\clr{1-\lr{-1}^m e^{-|\bm{k}_{\parallel}|d}} e^{i\bm{k}_{\parallel}\cdot\bm{\rho}}e^{ik_{zm} z}
  \label{F_sub-1} 
\end{align}

\bibliographystyle{apsrev4-2}

\begin{thebibliography}{21}%
\makeatletter
\providecommand \@ifxundefined [1]{%
 \@ifx{#1\undefined}
}%
\providecommand \@ifnum [1]{%
 \ifnum #1\expandafter \@firstoftwo
 \else \expandafter \@secondoftwo
 \fi
}%
\providecommand \@ifx [1]{%
 \ifx #1\expandafter \@firstoftwo
 \else \expandafter \@secondoftwo
 \fi
}%
\providecommand \natexlab [1]{#1}%
\providecommand \enquote  [1]{``#1''}%
\providecommand \bibnamefont  [1]{#1}%
\providecommand \bibfnamefont [1]{#1}%
\providecommand \citenamefont [1]{#1}%
\providecommand \href@noop [0]{\@secondoftwo}%
\providecommand \href [0]{\begingroup \@sanitize@url \@href}%
\providecommand \@href[1]{\@@startlink{#1}\@@href}%
\providecommand \@@href[1]{\endgroup#1\@@endlink}%
\providecommand \@sanitize@url [0]{\catcode `\\12\catcode `\$12\catcode `\&12\catcode `\#12\catcode `\^12\catcode `\_12\catcode `\%12\relax}%
\providecommand \@@startlink[1]{}%
\providecommand \@@endlink[0]{}%
\providecommand \url  [0]{\begingroup\@sanitize@url \@url }%
\providecommand \@url [1]{\endgroup\@href {#1}{\urlprefix }}%
\providecommand \urlprefix  [0]{URL }%
\providecommand \Eprint [0]{\href }%
\providecommand \doibase [0]{https://doi.org/}%
\providecommand \selectlanguage [0]{\@gobble}%
\providecommand \bibinfo  [0]{\@secondoftwo}%
\providecommand \bibfield  [0]{\@secondoftwo}%
\providecommand \translation [1]{[#1]}%
\providecommand \BibitemOpen [0]{}%
\providecommand \bibitemStop [0]{}%
\providecommand \bibitemNoStop [0]{.\EOS\space}%
\providecommand \EOS [0]{\spacefactor3000\relax}%
\providecommand \BibitemShut  [1]{\csname bibitem#1\endcsname}%
\let\auto@bib@innerbib\@empty
\bibitem [{\citenamefont {Brongersma}\ \emph {et~al.}(2015)\citenamefont {Brongersma}, \citenamefont {Halas},\ and\ \citenamefont {Nordlander}}]{brongersma2015plasmon}%
  \BibitemOpen
  \bibfield  {author} {\bibinfo {author} {\bibfnamefont {M.~L.}\ \bibnamefont {Brongersma}}, \bibinfo {author} {\bibfnamefont {N.~J.}\ \bibnamefont {Halas}},\ and\ \bibinfo {author} {\bibfnamefont {P.}~\bibnamefont {Nordlander}},\ }\href@noop {} {\bibfield  {journal} {\bibinfo  {journal} {Nature nanotechnology}\ }\textbf {\bibinfo {volume} {10}},\ \bibinfo {pages} {25} (\bibinfo {year} {2015})}\BibitemShut {NoStop}%
\bibitem [{\citenamefont {Li}\ \emph {et~al.}(2015)\citenamefont {Li}, \citenamefont {Coppens}, \citenamefont {Besteiro}, \citenamefont {Wang}, \citenamefont {Govorov},\ and\ \citenamefont {Valentine}}]{li2015circularly}%
  \BibitemOpen
  \bibfield  {author} {\bibinfo {author} {\bibfnamefont {W.}~\bibnamefont {Li}}, \bibinfo {author} {\bibfnamefont {Z.~J.}\ \bibnamefont {Coppens}}, \bibinfo {author} {\bibfnamefont {L.~V.}\ \bibnamefont {Besteiro}}, \bibinfo {author} {\bibfnamefont {W.}~\bibnamefont {Wang}}, \bibinfo {author} {\bibfnamefont {A.~O.}\ \bibnamefont {Govorov}},\ and\ \bibinfo {author} {\bibfnamefont {J.}~\bibnamefont {Valentine}},\ }\href@noop {} {\bibfield  {journal} {\bibinfo  {journal} {Nature communications}\ }\textbf {\bibinfo {volume} {6}},\ \bibinfo {pages} {8379} (\bibinfo {year} {2015})}\BibitemShut {NoStop}%
\bibitem [{\citenamefont {Ueno}\ \emph {et~al.}(2016)\citenamefont {Ueno}, \citenamefont {Oshikiri},\ and\ \citenamefont {Misawa}}]{ueno2016plasmon}%
  \BibitemOpen
  \bibfield  {author} {\bibinfo {author} {\bibfnamefont {K.}~\bibnamefont {Ueno}}, \bibinfo {author} {\bibfnamefont {T.}~\bibnamefont {Oshikiri}},\ and\ \bibinfo {author} {\bibfnamefont {H.}~\bibnamefont {Misawa}},\ }\href@noop {} {\bibfield  {journal} {\bibinfo  {journal} {ChemPhysChem}\ }\textbf {\bibinfo {volume} {17}},\ \bibinfo {pages} {199} (\bibinfo {year} {2016})}\BibitemShut {NoStop}%
\bibitem [{\citenamefont {Sundararaman}\ \emph {et~al.}(2014)\citenamefont {Sundararaman}, \citenamefont {Narang}, \citenamefont {Jermyn}, \citenamefont {Goddard~III},\ and\ \citenamefont {Atwater}}]{sundararaman2014theoretical}%
  \BibitemOpen
  \bibfield  {author} {\bibinfo {author} {\bibfnamefont {R.}~\bibnamefont {Sundararaman}}, \bibinfo {author} {\bibfnamefont {P.}~\bibnamefont {Narang}}, \bibinfo {author} {\bibfnamefont {A.~S.}\ \bibnamefont {Jermyn}}, \bibinfo {author} {\bibfnamefont {W.~A.}\ \bibnamefont {Goddard~III}},\ and\ \bibinfo {author} {\bibfnamefont {H.~A.}\ \bibnamefont {Atwater}},\ }\href@noop {} {\bibfield  {journal} {\bibinfo  {journal} {Nature communications}\ }\textbf {\bibinfo {volume} {5}},\ \bibinfo {pages} {5788} (\bibinfo {year} {2014})}\BibitemShut {NoStop}%
\bibitem [{\citenamefont {Manjavacas}\ \emph {et~al.}(2014)\citenamefont {Manjavacas}, \citenamefont {Liu}, \citenamefont {Kulkarni},\ and\ \citenamefont {Nordlander}}]{manjavacas2014plasmon}%
  \BibitemOpen
  \bibfield  {author} {\bibinfo {author} {\bibfnamefont {A.}~\bibnamefont {Manjavacas}}, \bibinfo {author} {\bibfnamefont {J.~G.}\ \bibnamefont {Liu}}, \bibinfo {author} {\bibfnamefont {V.}~\bibnamefont {Kulkarni}},\ and\ \bibinfo {author} {\bibfnamefont {P.}~\bibnamefont {Nordlander}},\ }\href@noop {} {\bibfield  {journal} {\bibinfo  {journal} {ACS nano}\ }\textbf {\bibinfo {volume} {8}},\ \bibinfo {pages} {7630} (\bibinfo {year} {2014})}\BibitemShut {NoStop}%
\bibitem [{\citenamefont {Besteiro}\ \emph {et~al.}(2017)\citenamefont {Besteiro}, \citenamefont {Kong}, \citenamefont {Wang}, \citenamefont {Hartland},\ and\ \citenamefont {Govorov}}]{besteiro2017understanding}%
  \BibitemOpen
  \bibfield  {author} {\bibinfo {author} {\bibfnamefont {L.~V.}\ \bibnamefont {Besteiro}}, \bibinfo {author} {\bibfnamefont {X.-T.}\ \bibnamefont {Kong}}, \bibinfo {author} {\bibfnamefont {Z.}~\bibnamefont {Wang}}, \bibinfo {author} {\bibfnamefont {G.}~\bibnamefont {Hartland}},\ and\ \bibinfo {author} {\bibfnamefont {A.~O.}\ \bibnamefont {Govorov}},\ }\href@noop {} {\bibfield  {journal} {\bibinfo  {journal} {Acs Photonics}\ }\textbf {\bibinfo {volume} {4}},\ \bibinfo {pages} {2759} (\bibinfo {year} {2017})}\BibitemShut {NoStop}%
\bibitem [{\citenamefont {Zhang}(2021)}]{zhang2021theory}%
  \BibitemOpen
  \bibfield  {author} {\bibinfo {author} {\bibfnamefont {Y.}~\bibnamefont {Zhang}},\ }\href@noop {} {\bibfield  {journal} {\bibinfo  {journal} {The Journal of Physical Chemistry A}\ }\textbf {\bibinfo {volume} {125}},\ \bibinfo {pages} {9201} (\bibinfo {year} {2021})}\BibitemShut {NoStop}%
\bibitem [{\citenamefont {Shi}\ \emph {et~al.}(2018)\citenamefont {Shi}, \citenamefont {Ueno}, \citenamefont {Oshikiri}, \citenamefont {Sun}, \citenamefont {Sasaki},\ and\ \citenamefont {Misawa}}]{shi2018enhanced}%
  \BibitemOpen
  \bibfield  {author} {\bibinfo {author} {\bibfnamefont {X.}~\bibnamefont {Shi}}, \bibinfo {author} {\bibfnamefont {K.}~\bibnamefont {Ueno}}, \bibinfo {author} {\bibfnamefont {T.}~\bibnamefont {Oshikiri}}, \bibinfo {author} {\bibfnamefont {Q.}~\bibnamefont {Sun}}, \bibinfo {author} {\bibfnamefont {K.}~\bibnamefont {Sasaki}},\ and\ \bibinfo {author} {\bibfnamefont {H.}~\bibnamefont {Misawa}},\ }\href@noop {} {\bibfield  {journal} {\bibinfo  {journal} {Nature nanotechnology}\ }\textbf {\bibinfo {volume} {13}},\ \bibinfo {pages} {953} (\bibinfo {year} {2018})}\BibitemShut {NoStop}%
\bibitem [{\citenamefont {Bohm}\ and\ \citenamefont {Pines}(1951)}]{bohm1951collective1}%
  \BibitemOpen
  \bibfield  {author} {\bibinfo {author} {\bibfnamefont {D.}~\bibnamefont {Bohm}}\ and\ \bibinfo {author} {\bibfnamefont {D.}~\bibnamefont {Pines}},\ }\href@noop {} {\bibfield  {journal} {\bibinfo  {journal} {Physical Review}\ }\textbf {\bibinfo {volume} {82}},\ \bibinfo {pages} {625} (\bibinfo {year} {1951})}\BibitemShut {NoStop}%
\bibitem [{\citenamefont {Pines}\ and\ \citenamefont {Bohm}(1952)}]{pines1952collective2}%
  \BibitemOpen
  \bibfield  {author} {\bibinfo {author} {\bibfnamefont {D.}~\bibnamefont {Pines}}\ and\ \bibinfo {author} {\bibfnamefont {D.}~\bibnamefont {Bohm}},\ }\href@noop {} {\bibfield  {journal} {\bibinfo  {journal} {Physical Review}\ }\textbf {\bibinfo {volume} {85}},\ \bibinfo {pages} {338} (\bibinfo {year} {1952})}\BibitemShut {NoStop}%
\bibitem [{\citenamefont {Bohm}\ and\ \citenamefont {Pines}(1953)}]{bohm1953collective3}%
  \BibitemOpen
  \bibfield  {author} {\bibinfo {author} {\bibfnamefont {D.}~\bibnamefont {Bohm}}\ and\ \bibinfo {author} {\bibfnamefont {D.}~\bibnamefont {Pines}},\ }\href@noop {} {\bibfield  {journal} {\bibinfo  {journal} {Physical Review}\ }\textbf {\bibinfo {volume} {92}},\ \bibinfo {pages} {609} (\bibinfo {year} {1953})}\BibitemShut {NoStop}%
\bibitem [{\citenamefont {Pines}(1953)}]{pines1953collective4}%
  \BibitemOpen
  \bibfield  {author} {\bibinfo {author} {\bibfnamefont {D.}~\bibnamefont {Pines}},\ }\href@noop {} {\bibfield  {journal} {\bibinfo  {journal} {Physical Review}\ }\textbf {\bibinfo {volume} {92}},\ \bibinfo {pages} {626} (\bibinfo {year} {1953})}\BibitemShut {NoStop}%
\bibitem [{\citenamefont {Yamada}(1960)}]{yamada1960mechanism}%
  \BibitemOpen
  \bibfield  {author} {\bibinfo {author} {\bibfnamefont {K.}~\bibnamefont {Yamada}},\ }\href@noop {} {\bibfield  {journal} {\bibinfo  {journal} {Kaku Yugo Kenkyu}\ }\textbf {\bibinfo {volume} {5}},\ \bibinfo {pages} {633} (\bibinfo {year} {1960})}\BibitemShut {NoStop}%
\bibitem [{\citenamefont {Economou}(1969)}]{economou1969surface}%
  \BibitemOpen
  \bibfield  {author} {\bibinfo {author} {\bibfnamefont {E.}~\bibnamefont {Economou}},\ }\href@noop {} {\bibfield  {journal} {\bibinfo  {journal} {Physical review}\ }\textbf {\bibinfo {volume} {182}},\ \bibinfo {pages} {539} (\bibinfo {year} {1969})}\BibitemShut {NoStop}%
\bibitem [{\citenamefont {Fetter}\ and\ \citenamefont {Walecka}(2012)}]{fetter2012quantum}%
  \BibitemOpen
  \bibfield  {author} {\bibinfo {author} {\bibfnamefont {A.~L.}\ \bibnamefont {Fetter}}\ and\ \bibinfo {author} {\bibfnamefont {J.~D.}\ \bibnamefont {Walecka}},\ }\href@noop {} {\emph {\bibinfo {title} {Quantum theory of many-particle systems}}}\ (\bibinfo  {publisher} {Courier Corporation},\ \bibinfo {year} {2012})\BibitemShut {NoStop}%
\bibitem [{\citenamefont {Ferrell}(1958)}]{ferrell1958predicted}%
  \BibitemOpen
  \bibfield  {author} {\bibinfo {author} {\bibfnamefont {R.~A.}\ \bibnamefont {Ferrell}},\ }\href@noop {} {\bibfield  {journal} {\bibinfo  {journal} {Physical Review}\ }\textbf {\bibinfo {volume} {111}},\ \bibinfo {pages} {1214} (\bibinfo {year} {1958})}\BibitemShut {NoStop}%
\bibitem [{\citenamefont {Kubo}(1957)}]{kubo1957statistical}%
  \BibitemOpen
  \bibfield  {author} {\bibinfo {author} {\bibfnamefont {R.}~\bibnamefont {Kubo}},\ }\href@noop {} {\bibfield  {journal} {\bibinfo  {journal} {Journal of the physical society of Japan}\ }\textbf {\bibinfo {volume} {12}},\ \bibinfo {pages} {570} (\bibinfo {year} {1957})}\BibitemShut {NoStop}%
\bibitem [{\citenamefont {Cho}(2003)}]{cho2003optical}%
  \BibitemOpen
  \bibfield  {author} {\bibinfo {author} {\bibfnamefont {K.}~\bibnamefont {Cho}},\ }\href@noop {} {\emph {\bibinfo {title} {Optical response of nanostructures: microscopic nonlocal theory}}},\ Vol.\ \bibinfo {volume} {139}\ (\bibinfo  {publisher} {Springer Science \& Business Media},\ \bibinfo {year} {2003})\BibitemShut {NoStop}%
\bibitem [{\citenamefont {Chew}(1999)}]{chew1999waves}%
  \BibitemOpen
  \bibfield  {author} {\bibinfo {author} {\bibfnamefont {W.~C.}\ \bibnamefont {Chew}},\ }\href@noop {} {\emph {\bibinfo {title} {Waves and fields in inhomogenous media}}}\ (\bibinfo  {publisher} {John Wiley \& Sons},\ \bibinfo {year} {1999})\BibitemShut {NoStop}%
\bibitem [{\citenamefont {Fano}(1961)}]{fano1961effects}%
  \BibitemOpen
  \bibfield  {author} {\bibinfo {author} {\bibfnamefont {U.}~\bibnamefont {Fano}},\ }\href@noop {} {\bibfield  {journal} {\bibinfo  {journal} {Physical review}\ }\textbf {\bibinfo {volume} {124}},\ \bibinfo {pages} {1866} (\bibinfo {year} {1961})}\BibitemShut {NoStop}%
\bibitem [{\citenamefont {Kanazawa}(1961)}]{kanazawa1961plasma}%
  \BibitemOpen
  \bibfield  {author} {\bibinfo {author} {\bibfnamefont {H.}~\bibnamefont {Kanazawa}},\ }\href@noop {} {\bibfield  {journal} {\bibinfo  {journal} {Progress of Theoretical Physics}\ }\textbf {\bibinfo {volume} {26}},\ \bibinfo {pages} {851} (\bibinfo {year} {1961})}\BibitemShut {NoStop}%
\end{thebibliography}

\end{document}